\documentclass{article}
\pdfoutput=1 

\usepackage{arxiv}
\usepackage{amsmath}
\usepackage{amssymb}
\usepackage{bm}
\usepackage[utf8]{inputenc} %
\usepackage[T1]{fontenc}    %
\usepackage{hyperref}       %
\usepackage{url}            %
\usepackage{booktabs}       %
\usepackage{amsfonts}       %
\usepackage{nicefrac}       %
\usepackage{microtype}      %
\usepackage{cleveref}       %
\usepackage{lipsum}         %
\usepackage{graphicx}       
\usepackage{doi}
\usepackage{mathtools}
\usepackage{svg}
\usepackage{siunitx}
\usepackage{xcolor}
\usepackage{caption}
\usepackage[backend=biber,style=numeric-comp,maxbibnames = 6,sorting = none,giveninits = true,minbibnames = 6]{biblatex}
\DeclareFieldFormat{pages}{\mkcomprange{#1}}

\usepackage{subcaption}

\newcommand{\orange}[1]{\textcolor{orange}{#1}}
\newcommand{\tm}[1]{\textrm{#1}}
\usepackage{float}
\usepackage[version=4]{mhchem} %
\usepackage{appendix}
\usepackage[flushleft]{threeparttable}
\title{Scale-bridging within a complex model hierarchy for investigation of a metal-fueled circular
energy economy by use of Bayesian model calibration with model error quantification}

\usepackage{authblk}
\author[a]{L. Gossel\thanks{Corresponding author\\Email address: gossel@mma.tu-darmstadt.de (L. Gossel)}}
\author[b]{E. Corbean}
\author[c]{S. Dübal}
\author[a]{P. Brand}
\author[a]{M. Fricke}
\author[d]{H. Nicolai}
\author[d]{C. Hasse}
\author[e]{S. Hartl}
\author[b]{S. Ulbrich}
\author[a]{D.~Bothe}

\affil[a]{Institute for Mathematical Modeling and Analysis, Technical University of Darmstadt, Peter-Grünberg-Straße 10, Darmstadt, 64287, Germany}
\affil[b]{Nonlinear optimization group, Department of Mathematics, Technical University of Darmstadt, Dolivostraße 15, Darmstadt, 64293, Germany}
\affil[c]{Department of Mechanical and Plastics Engineering, University of Applied Sciences Darmstadt, Holzhofallee 36b, Darmstadt, 64295, Germany}
\affil[d]{Institute for Simulation of Reactive Thermo-Fluid Systems, Technical University of Darmstadt, Otto-Berndt-Straße 2, Darmstadt, 64287, Germany}
\affil[e]{Faculty Mechanical and Systems Engineering, University of Applied Sciences Esslingen, Kanalstraße 33, Esslingen, 73728, Germany}
\renewcommand{\headeright}{}
\renewcommand{\undertitle}{}
\renewcommand{\shorttitle}{}

\begin{document}
\maketitle

\begin{abstract}

Metal energy carriers recently gained growing interest in research as a promising storage and transport material for renewable electricity. Within the development of a metal-fueled circular energy economy, research involves a model hierarchy spanning from micro to macro scales, making the transfer of information among different levels of complexity a crucial task for the implementation of the new technology. Chemical reactor networks (CRNs) are models of reduced complexity and a promising approach to accomplish the scale-bridging task. This holds if valid information from CRNs can be obtained on a much denser set of operating conditions than available from experiments and elaborated simulation methods like Computational Fluid Dynamics (CFD). An approach for CRN calibration from recent literature, including model error quantification, is further developed to construct a CRN model of a laboratory reactor for flash ironmaking, using data from the literature. By introducing a meta model of a CRN parameter, a simple CRN model on an extended set of operating conditions has successfully been calibrated. This way, the employed coupled calibration and uncertainty quantification framework has proven promising for the task of scale-bridging in the model hierarchy under investigation.
\end{abstract}

\section*{Abbreviations}
\begin{tabular}{ll}
  ABC   & Approximate Bayesian Computation\\
  a.u. & arbitrary units\\
  CFD   & Computational Fluid Dynamics  \\
  CRN   & Chemical Reactor Network\\
  MAP   & Maximum A Posteriori\\
  OP    & Operation Point \\
  PCE   & Polynomial Chaos Expansion \\
  PDF   & Probability Density Function \\
  PFR   & Plug Flow Reactor \\
  PSR   & Perfectly Stirred Reactor \\
\end{tabular}

\section*{Notation}
\begin{tabular}{ll}
  $\boldsymbol{\alpha}$ & Parameter vector for parametrization of model error\\
    $\boldsymbol{\Tilde{\alpha}}$ & $\boldsymbol{\Tilde{\alpha}}$ = ($\boldsymbol{\lambda}$,$\boldsymbol{\alpha}$)\\
    $B$ & Image space of $g$ and $f$, $B\subset \mathbb{R}$, $B=g(S) = f(S \times \mathbb{R}^d)$\\
  $D$   & Data set used for calibration, generated from $g$ \\
  $d$  & Dimension of model parameter vector $\lambda$\\
  $\delta$ & (Unknown) discrepancy between $g$ and $f$\\
  $\epsilon$ & Data error occuring in the evaluation of $g$\\
  $f$   & Low-fidelity parametrized model to be calibrated\\
  $g$   & High-fidelity model $g$ used for calibration of $f$\\
  $\eta$ & Tolerance parameter in the ABC likelihood construction\\
  $\boldsymbol{\lambda}$ & Vector of parameters to be calibrated in the model $f$\\
    $\boldsymbol{\Lambda}$ & Random vector representing the  model parameters augmented with model error\\
    &($\boldsymbol{\Lambda}(\boldsymbol{\alpha},\boldsymbol{\xi}) = \boldsymbol{\lambda} + \boldsymbol{\delta}(\boldsymbol{\alpha},\boldsymbol{\xi})$)\\
    $\mu$& Mean of a random quantity\\
  $N$ & Dimensionality of design space\\
  $\pi_Q(\cdot)$ & Probability density function with respect to a random quantity Q\\
  $S$   & Design space, $D \subsetneq S \times B \subset \mathbb{R}^N\times B$\\
  $\sigma$ & Standard deviation of a random quantity\\
  $\mathbf{x}$ & Vector of observables/oeprating conditions ($\mathbf{x} \in S$)\\
  $\boldsymbol{\xi}$ & Vector of independent identically distributed random variables\\
  $y$ & Model output observable ($y \in B)$
\end{tabular}

\clearpage
\section*{Nomenclature}

\begin{table}[H]
    \centering
\begin{tabular}{lll}
Name&Meaning throughout this work&Example\\
\hline
     Operating condition&  A set of parameters that can be & Gas and solid mass flows provided to a reactor \\
     &adjusted in experiments& \\
     & and is not dependent on a model.& \\
     &Serves as input to models. & \\
     \hline
     Operation point& A specific, labeled set of operation & see Table~\ref{TC_OPTable}\\
     &conditions used within& \\
     & the study, mostly accompanied & \\
     &with a measurement result& \\
     \hline
     Model parameter& Parameters specific to a certain model. & CRN: reactor temperatures, volumes,....\\
     & The calibration procedures presented& \\
     & in this work are for calibration& \\
     & of (a subset of) model parameters& \\
     \hline
     Reactor&Is used ambiguously as term &(1) Laboratory reactor for flash ironmaking\\
     &for a full chemical reactor (1) & \\
      &or the modeling blocks & (2) PFR and PSR models (see Section~\ref{CRN})\\
     &within a CRN model (2) & \\
     \hline
     Design space& Subset of the space of \textbf{operating } & Figure~\ref{TC_DesignSpace}\\
     &\textbf{conditions} for which a valid (CRN)  & \\
     &model shall be calibrated.& \\
      \hline
\end{tabular}
\end{table}

\section{Introduction \label{Introduction}}

Metal fuels and especially iron are recently gaining attention as carbon-neutral chemical energy carriers for storing and transporting renewable electricity \cite{julien2017, bergthorson2015, janicka2023} due to advantageous 
properties of the considered metals, such as abundance and volumetric energy density. In a circular process, energy from renewable energy sources is stored in form of iron through reduction of iron oxides using green hydrogen as reducing agent. The iron particles can then be transported to locations with high energy demand, where energy is released in form of heat through combustion with oxygen, producing iron oxide particles that can be returned to the reduction site to close the cycle.

Until ready for deployment, the investigation of such a novel concept relies on various different scales ranging from fundamental research on the micro- and nanoscale \cite{Mich2023,Buchheiser2023,Kuhn2022,Li2022}, via mesoscale laboratory and pilot systems \cite{Krenn2023,Fedoryk2023,Fradet2023,Duebal2024}, up to global scale investigations such as techno-economic and lifecycle assessments \cite{Neumann2022,Neumann2023,Neumann2024} or political and economical considerations \cite{Bruch2023,Ott2022,Plank2023}. Information transfer within this model hierarchy on a vast range of scales is crucial for proper model analyses and for ensuring the technology is ready for application within the urgent time frame of the energy transition. 
This involves several aspects, such as enabling the use of insights from the mirco scale models even in the largest scale optimization problems, and providing access to the full design space of possible conditions and configurations for these models, while maintaining valid model responses. For the scale-bridging task defined as such, mesoscale applications are of particular importance, as they naturally reduce the scale gap between the micro and macro level. 
There, chemical reactors from laboratory to pilot industrial scales are investigated experimentally and with modeling approaches and simulations. 

In the context of reactor modeling, Computational Fluid Dynamics (CFD) simulations are typically used for detailed reactor simulations that enable deep understanding of the physical phenomena in the reactor such as global and local fluid dynamics, turbulence-chemistry interactions and flame structures \cite{Hasse2021,Nicolai2022}. 
However, performing CFD simulations is computationally costly. Experiments are costly, too, and usually take a long time to be accomplished. Experiments and CFD simulations yield crucial information and understandings of the overall reactor behavior. 
To a certain extent, these methods also yield information on the chemical conversion of species, 
the latter however being restricted to certain species, depending on the case, and on the available measurement techniques or computational resources\cite{Hasse2021,Nicolai2022}. 

Scale- and complexity-reduced models try to find a trade-off between the (computational) cost and the considered level of detail and hence represent a bridge between detailed expensive simulation models and the need for readily accessible information on the global scale. Chemical reactor networks (CRNs) \cite{Trespi2021} are an instance for scale-reduced models that are based on CFD or experimental results, or both. CRNs are graph-like structures composed of modeling blocks, which employ strongly simplified flow models, while enabling the incorporation of reaction mechanisms that are partially even more complex than those tractable in CFD simulations \cite{Trespi2021,Duebal2024}. 
The computational effort of solving CRNs depends on the applied reaction mechanism and the number of reactor compartments, their structure within the network as well as the underlying reactor description. Nevertheless, using CRNs leads to a great reduction of computation time and resources in comparison to CFD models. CRN solutions can be computed on personal computers  in less than seconds to a few hours, while CFD computations usually take days to weeks on high performance clusters \cite{Trespi2021}. 

However, the construction of CRNs is usually based on CFD or experimental results \cite{Trespi2021,Duebal2024}. In order to be used as a scale-bridging tool in large-scale mathematical and thermodynamic models, data is needed on a set of operating conditions that is much larger than the available data set from CFD or experiments, 
which implies that CRNs need to be validly constructed and solved for those conditions where no CFD or experimental data exists. A promising approach to analyse and exploit the predictive capacity of CRNs has recently been introduced by Savarese \textit{et al.} \cite{Savarese2023b}, who exemplified the calibration of a CRN model to a set of CFD data points within a Bayesian model-to-model-calibration framework, including model error quantification as proposed in \cite{Sargsyan2015,Sargsyan2019}, for an ammonia-fueled micro-turbine. 

The objective of the present work is to transfer this approach to the present setting of scale-bridging within a metal-fueled energy circular economy and to preliminarily assess the usability of the calibration framework within this scope. 
The study concentrates on the reduction site of the energy cycle. Flash ironmaking \cite{Sohn2023} is a novel technology for iron oxide reduction that has several advantages over conventional iron production concepts like blast furnace processes \cite{Sohn2023}, and is also a promising candidate when considering iron as energy carrier within a circular economy \cite{Neumann2022,Neumann2024}. In an extensive study, Sohn and co-workers have studied this process with applications on various scales, from kinetic development in laminar flow reactors, via laboratory and pilot experimental applications, to CFD simulations from the laboratory to industrial scale \cite{Chen2015,Fan2016,Fan2016a,Fan2017,Abdelghany2019,Abdelghany2020,Sohn2021,Sohn2021a,Sohn2023}. Within the scope of the present work, a CRN model is developed 
for modeling the laboratory reactor by Sohn, using the experimental and simulation data summarized in \cite{Sohn2023} for CRN construction, calibration and validation. The use of literature data from a former study comes along with the challenge of using experimental data for which operation conditions corresponding to the data being spread irregularly and unevenly in the design space. As such, the chosen application is a valid instance for assessing the actual usability of CRN models as a bridging element in the present model and scales hierarchy and thus, its quality for promoting the faster deployment of the innovative energy circular economy behind. 

The paper is structured as follows. Section \ref{Theoretical Background} summarizes the theoretical concept of chemical reactor network modeling, and model-to-model calibration with embedded model error in the framework of Bayesian inference, based on the work by Sargsyan \textit{et al.} \cite{Sargsyan2015,Sargsyan2019}. Section \ref{Methodology} describes the developed workflow used to employ the theoretical concepts to the use case, which is similar to the workflow introduced by Savarese \textit{et al.} \cite{Savarese2023b}. Further details on the testcase of CRN calibration for iron oxide reduction will be provided in Section \ref{Testcase}. The CRN construction and obtained calibration results will be presented and discussed in Section \ref{Results} and the results will be 
related to the objectives and 
the scope of this work in Section~\ref{Discussion}. Lastly, Section \ref{Conclusion + Outlook} concludes this work with final remarks and an outlook on future work.  

\section{Theory}\label{Theoretical Background}

The aim of the present work is to calibrate a CRN model on a given set of operating conditions, the design space. In order to assess the validity of the obtained calibrated model, it is crucial to quantify uncertainties of the model output and, where possible, investigate and mitigate their origin. Because of this, we employ the embedded model error approach suggested  by Sargsyan \textit{et al.}\cite{Sargsyan2015}, which has already been exemplified for CRNs by Savarese \textit{et al.}\cite{Savarese2023b}.

In the following sections, we will introduce the key ideas of that approach relevant to understanding our application and also the calibration procedure, mostly following the work of Sargsyan et al. \cite{Sargsyan2015,Sargsyan2019}. Before that, we will shortly introduce the concept of chemical reactor networks. 

\subsection{Chemical Reactor Network Modeling \label{CRN}}
Chemical reactor networks (CRNs) are graph-like mathematical models of a chemical reactor, built of modeling units representing spatial regions of the reactor, which are interconnected by streams holding the information about the mass flows between the compartments. The unit boundaries are determined by the physical properties within the reactor like temperature distribution and flow structure, leading to the definition of functional compartments \cite{Trespi2021}. CRNs are modeling the stady-state behaviour of the reactors, i.e. the total time derivative of the system vanishes. 
\begin{figure}[H]
\centering
\includegraphics[width=0.8\textwidth]{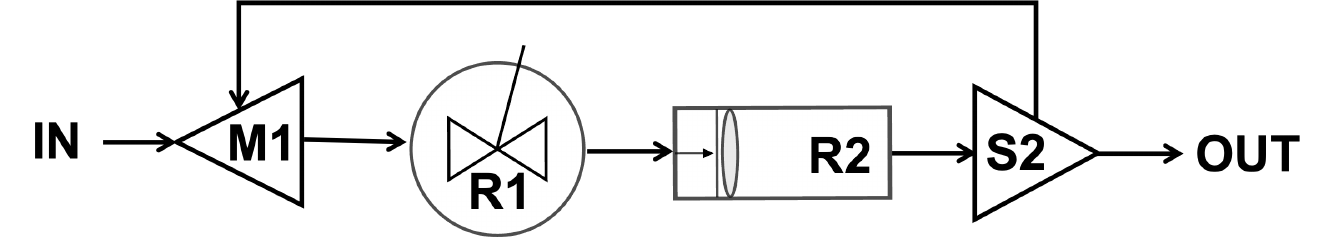}
\caption{Generic chemical reactor network including ideal reactors, mixing and splitting units. Adapted from \cite{Trespi2021}.} 
\label{fig:GenericCRN}
\end{figure}
The main units of a CRN are ideal steady-state reactor models that combine a simplified flow assumption with a detailed description of chemical reactions. 
A generic reactor network is depicted in Figure~\ref{fig:GenericCRN}. 
Regions of strong mixing are approximated by the perfectly stirred reactor (R1), a 0D model, which assumes infinitely fast and perfect mixing of the species. %
The chemical system is defined by the density $\rho$ of the mixture, the molecular weight $\mathrm{MW_i}$, the formation rate $\Dot{\omega}_i$ and the mass fraction $Y_i$ of the reactants and the residence time $\tau$ of the reactor \cite{Trespi2021}: 
\begin{align}
    \frac{dY_i}{dt} = \frac{Y_i^{in}-Y_i}{\tau} + \frac{\mathrm{MW}_i\Dot{\omega}_i}{\rho}.
    \label{PSR_massfrac_equation}
\end{align}
Here, \textit{i} is indexing the single species. 
The 1D plug flow reactor model (R2) considers straight flow patterns, assumes perfect radial mixing and neglects mixing in the axial direction. Thus, the species conservation equations along the reactor coordinate $z$ are formulated as follows \cite{Trespi2021}:
\begin{align}
    \frac{dY_i}{dz} = \frac{\mathrm{MW}_i\Dot{\omega}_i}{\rho v}
     \label{PFR_massfrac_equation}
\end{align}
with the fluid velocity $v$. Together with mixing (M1) and splitting units (S1) to model the distribution of the flow between the different reactors, the CRN structure is fully determined. 
The solution of a CRN model is a vector of mass fractions for the given species and every stream and reactor of the CRN. Here, however, usually the outlet stream of the complete reactor, i.e. the stream leaving the network, is of interest. 

The modeling parameters of a CRN are the overall structure and number of modeling blocks and several parameters within each modeling block, like the temperature, volumes or residence times. The 
modeling blocks are assumed to be isothermal throughout this work, which is often the case for CRNs. 
The enthalpy equations having the most influence on temperature are usually solved within the CFD computation and the resulting temperature fields are then used for the CRN construction. 
A subset of the CRN model parameters will be calibrated within this work, using the calibration framework introduced in the next sections. 

\subsection{Model-to-Model-calibration with embedded model error \label{TB_Model}}
Consider a  set $S \subset \mathbb{R}^N,\ N\in\mathbb{N}$ of physical observables, which we will refer to as the design space. In our case, these observables are operating conditions like mass flows or pressure, on which the CRN depends. The design space defines the bounds within which a valid model is to be calibrated. 

Consider a model $g\colon S \rightarrow B \subset \mathbb{R},\ \mathbf{x}\longrightarrow y$, mapping the design space to some observable $y$, which is throughout this work assumed to be scalar. 

This model is referred to as \textit{high-fidelity}, or even \textit{truth} model \cite{Sargsyan2015}, which essentially means that $g$ is assumed to have negligible model error, compared to another model $f$ that shall be calibrated on $S$. On the other hand, $g$ is in general expensive to be evaluated such that only a finite set of data generated from $g$ is available, which is denoted by $D = \{\left(\mathbf{x}^{(i)}, y^{(i)}\right)\}_{i=1}^L$, where $L$ is the number of evaluations of $g$ and $\mathbf{x}^{(i)} \in S$ is the $i$-th operation condition for which $g$ is evaluated. In general, there can be a data error while evaluating $g$, leading to

\begin{equation}
    y^{(i)} = g(\mathbf{x}^{(i)}) + \epsilon^{(i)},
\end{equation}

where $\epsilon^{(i)}$ is the data error associated to the evaluation of $y^{(i)}$. In the context of chemical reactor networks, the data set $D$ is usually obtained from experiments or CFD simulations. 

The model to be calibrated is a parametrized model $f\colon S\times\mathbb{R}^d \rightarrow B\subset \mathbb{R},\ (\mathbf{x},\boldsymbol{\lambda})\longrightarrow y$, where $\boldsymbol{\lambda} = (\lambda_1,...,\lambda_d)$ is the vector of parameters being specific to the model $f$. 
In the CRN context this can for instance be reactor temperatures or volumes. The vector $\boldsymbol{\lambda}$ provides the degrees of freedom for the calibration of $f$ to the data set $D$. The model $f$ is referred to as \textit{low-fidelity} model as it has a non-negligible model error, as will be introduced next. Note that $f$ is assumed to be surjective and, more strictly, it is assumed that for every $y\in B$ and every $\mathbf{x}\in S$ there is a $\bm{\lambda}\in \mathbb{R}^d$ such that $f(\mathbf{x},\bm{\lambda}) = y$. \\

In addition to the measurement noise represented by $\epsilon^{(i)}$, there is a model discrepancy between the high-fidelity model $g(\mathbf{x})$ and the lower fidelity model $f(\mathbf{x}, \bm{\lambda})$, defined as $\delta(\mathbf{x}, \bm{\lambda}) = g(\mathbf{x}) - f(\mathbf{x}, \bm{\lambda})$. This results in the following expression for the relation of the lower fidelity model $f$ to the measurements $(\mathbf{x}^{(i)}, y^{(i)})$: 
\begin{align}
    y^{(i)} = f(\mathbf{x}^{(i)}, \bm{\lambda}) + \delta(\mathbf{x}^{(i)}, \bm{\lambda}) + \epsilon^{(i)}.
\end{align}
One way of dealing with this explicit formulation of the model error $\delta(\mathbf{x})$ as presented in \cite{Sargsyan2019} is to embed the model error within the model $f$. For this purpose, an additive term $\bm{\delta}$ is added to the model parameters $\bm{\lambda}$ in order to account for the model discrepancy, resulting in the term $f(\mathbf{x}, \bm{\lambda} + \bm{\delta})$. 
As we will shortly see, the additive term $\bm{\delta}$ will be represented by a polynomial chaos expansion parameterized by parameters $\bm{\alpha}$ and expressed in terms of a random vector $\bm{\xi}$, such that the model can be rewritten as 
\begin{align}
    y^{(i)} \approx f(\mathbf{x}^{(i)}, \bm{\lambda} + \bm{\delta}(\bm{\alpha}, \bm{\xi})) + \epsilon^{(i)}\ \text{for}\ i=1,\dots, L.
\end{align}
In the following, the parameters $\bm{\lambda}$ and $\bm{\alpha}$ are combined into $\bm{\Tilde{\alpha}} = (\bm{\lambda}, \bm{\alpha})$ and the random quantity representing the parameters $\bm{\lambda}$ and the additive uncertain model error $\bm{\delta}(\bm{\alpha}, \bm{\xi})$ is denoted by $\bm{\Lambda}(\bm{\Tilde{\alpha}}) = \bm{\lambda} + \bm{\delta}(\bm{\alpha}, \bm{\xi})$, resulting in 
\begin{align}
        y^{(i)} \approx f(\mathbf{x}^{(i)}, \bm{\Lambda}(\bm{\Tilde{\alpha}})) + \epsilon^{(i)}\ \text{for}\ i=1,\dots, L.
\end{align}

It is important to note that the original calibration for $\bm{\lambda}$ has been turned into a parameter estimation problem for $\bm{\Tilde{\alpha}}$ by explicitly embedding the model error. The result of the calibration is a PDF associated to a random quantity $\bm{\Lambda}$, which in turn yields a random prediction $f(\mathbf{x},\bm{\Lambda})$, the latter being characterized by a PDF, too. 
The random prediction $f(\mathbf{x},\bm{\Lambda})$ and its associated PDF can in the end be interpreted as an approximation of the probability distribution of the "true" prediction $g(\mathbf{x})$.

\subsection{Bayesian Inference for Model Calibration}
Bayesian and frequentist statistics represent two distinct approaches used for statistical inference, that differ in the way of interpreting probability and using observed data, leading to different inference concepts \cite{Bolstad2016}. While the frequentist approach views probabilities as the long-term relative frequency of occurrence of a certain event, in the Bayesian approach probabilities represent beliefs that are updated based on observations. \\
Bayesian inference incorporates prior knowledge on parameters $\bm{\tilde{\alpha}}$, i.e. it views $\bm{\tilde{\alpha}}$ as random vector that is distributed according to some probability distribution $\mathcal{P}(\bm{\tilde{\alpha}})$, and computes a posterior distribution $\mathcal{P}(\bm{\tilde{\alpha}} | D)$ of the parameters, given some observations $D$, using Bayes' Theorem 
\begin{align}
    \mathcal{P}(\bm{\tilde{\alpha}} | D) = \frac{\mathcal{P}(D | \bm{\tilde{\alpha}}) \cdot \mathcal{P}(\bm{\tilde{\alpha}})}{\mathcal{P}(D)}.
\end{align}
Here, $\mathcal{P}(D|\bm{\tilde{\alpha}})$ is the likelihood function that represents the probability of the observations given specific parameter values $\bm{\tilde{\alpha}}$, whereas $\mathcal{P}(D)$, sometimes referred to as evidence, is the probability of the observations. Note that $\mathcal{P}(D)$ is independent of the realization of $\bm{\tilde{\alpha}}$ and hence can be seen as a normalizing factor. \\
In the presented setting of model-to-model calibration, the goal is to determine parameters $\bm{\tilde{\alpha}}$ that best describe the observed data $D$. Thus, computing the posterior distribution $\mathcal{P}(\bm{\tilde{\alpha}} | D)$ provides the desired knowledge about $\bm{\tilde{\alpha}}$ in terms of the probability that, given the observations $D$, the model is parameterized by $\bm{\tilde{\alpha}}$. Hence the model-to-model calibration problem can be solved by performing Bayesian inference for the parameters $\bm{\tilde{\alpha}}$. \\
In the following, the general calibration procedure will be presented, including how the different components (e.g. prior, likelihood) are constructed in order to perform Bayesian inference for obtaining the posterior distribution.

\subsection{Calibration Procedure \label{TB_CaliP}}

In Table~\ref{Cali_Prod_Overview}, an overview on the calibration procedure is given. The order of some of the steps is interchangeable, however, for the purpose of orientation, 
the order has been chosen as presented here. In the present section, the theoretical background of the calibration steps will be explained in the order given in the table. 

\begin{table}[h!]
    \centering
    \begin{tabular}{ll}
    \hline 
 1.       &Parametrization of model error to obtain the parameter set $\bm{\tilde{\alpha}}$ \\
  2.       &Definition of priors \\
  3.       &Choose likelihood type (here: ABC) \\
    &3.1. ABC: Choose parameter $\eta$ \\
  4.& Sampling of $\bm{\tilde{\alpha}}$ with the Markov Chain Monte Carlo (MCMC) method\\
   5.& For each $\bm{\tilde{\alpha}}$: Calculation of likelihood. \\
      &5.1.  Model error embedding: create PCE representation of $f(\mathbf{x},\bm{\Lambda}(\bm{\tilde{\alpha}}))$\\
 & 5.2. ABC: Calculate mean and variance of $f(\mathbf{x},\bm{\Lambda}(\bm{\tilde{\alpha}}))$ (straight-forward with PCE representation created in 5.1.)\\ & 5.3. ABC: Calculate $L_{ABC}(\bm{\tilde{\alpha}})$ with Equation~\eqref{LikABC}\\
   6.& Obtain posterior value for $\bm{\tilde{\alpha}}$,  $\mathcal{P}(\bm{\tilde{\alpha}}|D) \propto \mathcal{P}(D|\bm{\tilde{\alpha}})\mathcal{P}(\bm{\tilde{\alpha}}) = L_{ABC}(\bm{\tilde{\alpha}})\mathcal{P}(\bm{\tilde{\alpha}})$ \\
      7.& Continue sampling of $\bm{\tilde{\alpha}}$ according to posterior value obtained.  \\
     & 7.1. Back to 5., until maximum MCMC step number is reached. \\
   \hline
    \end{tabular}
        \caption{Overview of the calibration procedure employed}
    \label{Cali_Prod_Overview}
\end{table}

\subsubsection{Polynomial chaos expansions of the parameters augmented with model error \label{PCE}} \label{PCE}

Within the Bayesian inference approach applied in this work, Polynomial Chaos Expansions (PCEs) have an important role for both, model error embedding, and enhancing the numerical tractability. 

Polynomial Chaos expansions are feasible representations of random quantities such as the model parameters augmented by model error, i.e. $\bm{\Lambda}(\bm{\Tilde{\alpha}})$. Consider a polynomial basis $\{\psi^{(i)}(\xi) |\ i \in \mathbb{N}_0\}$ with $\deg (\psi^{(i)}) = i$. Assume that $\xi$ is a random variable supported on $[-1,1]$ and that the polynomial basis is orthogonal with respect to the PDF of $\xi$ given by $\pi(\xi)$, that is
\begin{equation}
    \langle \psi^{(i)}, \psi^{(j)} \rangle := \int_{[-1,1]} \psi^{(i)}(\xi)\psi^{(j)}(\xi)\pi(\xi) d \xi = \delta_{ij}\langle \psi^{(i)}, \psi^{(i)} \rangle,
\end{equation}
where $\delta_{ij}$ is the Kronecker delta. 

Furthermore, consider a maximum polynomial order $p \in \mathbb{N}$, a set of multi-indices 
\begin{align}
K = \{\mathbf{k}=(k_1,...,k_d) \in \mathbb{N}^d \ |\ |\mathbf{k}| = \sum_{j=1}^d k_j \leq p \}
\end{align} and a set of coefficients $\{a_{\mathbf{k}} | \mathbf{k} \in K \}$. Let $\bm{\xi} = (\xi_1, \dots, \xi_d)$ be a vector of independent identically distributed variables.
Then a general finite-variance random quantity $\zeta$ can be represented as \cite{Sargsyan2015,Ghanem1991,Wiener1938}:

\begin{equation}
    \zeta = \sum_{\ell\in \mathbb{N}} a_{\mathbf{k}_\ell} \Psi_{\mathbf{k}_\ell}(\boldsymbol{\xi}), 
\end{equation}

where $\Psi_{\mathbf{k}}(\boldsymbol{\xi}) = \Pi_{j=1}^{d} \psi^{(k_j)}(\xi_j)$. 
In numerical practice, a truncation up to an order $p$ is used to approximate $\zeta$ \cite{Ghanem1991,Wiener1938,Debusschere2004} according to 

\begin{equation}
\zeta \approx \sum_{\ell=1}^{P} a_{\mathbf{k}_\ell} \Psi_{\mathbf{k}_\ell}(\boldsymbol{\xi}),
\label{TB_TruncatedPCE}
\end{equation}

involving $P = \frac{(p+d)!}{p!d!} = |K|$ terms. Note that each $\Psi_{\mathbf{k}}(\boldsymbol{\xi})$ is a product of the basis polynomials, where the maximum sum of all polynomial orders of the $d$ polynomials is $p$ and $\zeta$ is the sum of all possible $\Psi_{\mathbf{k}}(\boldsymbol{\xi})$ fulfilling this constraint, each weighted with a coefficient $a_{\mathbf{k}_\ell}$. 

This way, the $j$-th augmented model parameters $\Lambda_j (\bm{\tilde{\alpha}}, \boldsymbol{\xi})$ can be represented via Polynomial Chaos Expansion \cite{Sargsyan2019} as

\begin{equation}
    \Lambda_j (\boldsymbol{\tilde{\alpha}}, \boldsymbol{\xi}) = \lambda_j + \delta_j(\boldsymbol{\alpha}, \boldsymbol{\xi}) \approx \lambda_j + \sum_{\ell=1}^{P} \alpha _{\mathbf{k}_{\ell,j}} \Psi_{\mathbf{k}_\ell}(\boldsymbol{\xi}) = \Tilde{\alpha}_{0,j} + \sum_{\ell=1}^{P} \Tilde{\alpha} _{\mathbf{k}_{\ell,j}} \Psi_{\mathbf{k}_\ell}(\boldsymbol{\xi}). 
\end{equation}

In Section~\ref{TB_PredMomEstim}, the PCE approach will be further employed for creating surrogate models of $f(\mathbf{x},\bm{\Lambda}(\bm{\Tilde{\alpha}}))$, and in Section~\ref{Surrogate}, a deterministic approach with similar structure will be employed for creating surrogate models of $f(\mathbf{x},\bm{\lambda})$, since surrogate models usually have a higher numerical tractability than the model itself.  

\subsubsection{Definition of priors}

In order to calculate the posterior distribution $\mathcal{P}(\boldsymbol{\Tilde{\alpha}}|D)$, the prior $\mathcal{P}(\boldsymbol{\Tilde{\alpha}})$ needs to be defined. The prior contains the knowledge on the probability distribution of $\boldsymbol{\Tilde{\alpha}}$ \textit{before} the calibration. Throughout the present work, bounded uniform priors will be employed, i.e.

\begin{equation}
    \mathcal{P}(\Tilde{\alpha}_i) = \begin{cases}
      \textrm{const},   & a_i<\Tilde{\alpha}_i<b_i, \\
      0,   & \textrm{else},
    \end{cases}
\end{equation}
where $\Tilde{\alpha}_i$ is the $i$-th component of $\boldsymbol{\Tilde{\alpha}}$. The respective bounds $a_i$ and $b_i$ are determined by consideration of the physical knowledge on the parameters. This way, no additional prior knowledge is incorporated into the calibration, which is desired in the present work. 

\subsubsection{Calculation of likelihood \label{TB_Likelihood}}

The calculation of a proper likelihood $L_D(\boldsymbol{\Tilde{\alpha}}) = \mathcal{P}(D|\boldsymbol{\Tilde{\alpha}})$ is a central step in the inference procedure. Given a parametrization $\boldsymbol{\Tilde{\alpha}}$ of the random parameter set $\boldsymbol{\Lambda}(\boldsymbol{\Tilde{\alpha}})$, it yields the probability that the data set $D$ will be obtained when the model $f$ is evaluated at the operation conditions given in $\mathbf{x}$.

However, the latter statement can be interpreted in different ways, depending on how exactly the likelihood is calculated. This will be illustrated in the following. 
 
Sargsyan and coworkers provide a detailed overview of the different ways the likelihood can be modeled, including the features and shortcomings of the different approaches \cite{Sargsyan2015,Sargsyan2019}. This given, only a few central aspects will be summarized, especially focusing on the proper interpretation of the results that are obtained when choosing a likelihood type.  

The first important point is the following distinction:
Either it is required that there exists a realisation of the parameter vector $\boldsymbol{\Lambda}(\boldsymbol{\Tilde{\alpha}})$ 
such that $f(\mathbf{x},\boldsymbol{\Lambda}(\boldsymbol{\Tilde{\alpha}})) = g(\mathbf{x})$ for \textbf{every} $\mathbf{x} \in S$. Otherwise, it is sufficient that for each $\mathbf{x} \in S$, there is a realisation of $\boldsymbol{\Lambda}(\boldsymbol{\Tilde{\alpha}}(\mathbf{x}))$ such that $f(\mathbf{x},\boldsymbol{\Lambda}(\boldsymbol{\Tilde{\alpha}}(\mathbf{x})))=g(\mathbf{x})$, but here the value of $\boldsymbol{\Lambda}(\boldsymbol{\Tilde{\alpha}}(\mathbf{x}))$ can be different for every $\mathbf{x}$. This distinction is incorporated into the likelihood evaluation by either requiring that there is a realisation $\boldsymbol{\Lambda}(\boldsymbol{\Tilde{\alpha}})$ with nonzero probability density such that $f(\mathbf{x}^{(i)},\boldsymbol{\Lambda}(\boldsymbol{\Tilde{\alpha}}))=y^{(i)}$ for \textbf{every} $\left(\mathbf{x}^{(i)},y^{(i)}\right) \in D$ for $\boldsymbol{\Tilde{\alpha}}$ to have nonzero likelihood, or only requiring that for every $\left(\mathbf{x}^{(i)},y^{(i)}\right) \in D$, there is a realisation $\boldsymbol{\Lambda}(\boldsymbol{\Tilde{\alpha}}(\mathbf{x}))$ with nonzero probability density such that $f(\mathbf{x}^{(i)},\boldsymbol{\Lambda}(\boldsymbol{\Tilde{\alpha}}(\mathbf{x})))=y^{(i)}$, if $\boldsymbol{\Tilde{\alpha}}$ is to be assigned nonzero likelihood. 

In the first case, the $L$-dimensional PDF $\pi_{\mathbf{f}(\boldsymbol{\Tilde{\alpha}},\boldsymbol{\xi})}(\mathbf{\cdot})$ with $\mathbf{f}(\boldsymbol{\Tilde{\alpha}},\boldsymbol{\xi})\coloneqq \left(f(\mathbf{x}^{(1)},\boldsymbol{\Tilde{\alpha}},\boldsymbol{\xi})+\epsilon^{(1)},...,f(\mathbf{x}^{(L)},\boldsymbol{\Tilde{\alpha}},\boldsymbol{\xi})+\epsilon^{(L)}\right) $ has to be computed and evaluated at $\mathbf{y} \coloneqq \left(y^{(1)},...,y^{(L)}\right)$, with $(\mathbf{x}^{(i)},y^{(i)}) \in D$. This is referred to as the full likelihood by Sargsyan \textit{et al.} \cite{Sargsyan2015,Sargsyan2019} and, if no data error is considered ($\epsilon^{(i)} = 0\ \textrm{for}\ i=1,...,L$), it is in general zero for any $\boldsymbol{\Tilde{\alpha}}$, except the trivial case in which there exists a parameter vector $\boldsymbol{\Lambda}(\boldsymbol{\Tilde{\alpha}})$ such that $f(\cdot,\boldsymbol{\Lambda}(\boldsymbol{\Tilde{\alpha}})) = g(\cdot)$. If a data error is considered, it becomes finite, but the $L$-dimensional PDF remains computationally intractable in many cases \cite{Sargsyan2015,Sargsyan2019}. 

In the second case, the likelihood is calculated as the product of the marginal PDFs for each $\left(\mathbf{x}^{(i)},y^{(i)}\right) \in D$, that is \cite{Sargsyan2015,Sargsyan2019}
\begin{equation}
    L_D(\boldsymbol{\Tilde{\alpha}}) = \Pi_{i=1}^{L} \pi_{\left(f(\mathbf{x}^{(i)},\boldsymbol{\Tilde{\alpha}},\boldsymbol{\xi})+\epsilon^{(i)}\right)}(y^{(i)}). 
    \label{MarginalLikelihood}
\end{equation}

From a stochastic point of view, if a "full" likelihood is employed, the components of $\mathbf{f}$ are treated as dependent variables, whereas if the "marginal" likelihood is employed, they are considered independent. 

The marginal likelihood is in general nonzero for many samples of $\boldsymbol{\Tilde{\alpha}}$ and much more tractable from the computational point of view, as $L$ one-dimensional PDFs have to be calculated instead of one $L$-dimensional PDF. Still, even the marginal likelihood~\eqref{MarginalLikelihood} can be too demanding in terms of computational resources, which is why several moment-based approximations are suggested by Sargsyan \textit{et al.} \cite{Sargsyan2015,Sargsyan2019}, one of them being introduced below and employed for the computations that have been conducted for the present work. 

Before this, a short comment shall be made on the interpretation of results when a "full" or "marginal" likelihood type is used. Especially, if no data error is considered, the marginal likelihood is more than just an approximation of the full likelihood. In the end, it (approximately) yields what the model error approach aims for: 
A PDF over a random parameter vector $\boldsymbol{\Lambda}$, parametrized by a vector $\boldsymbol{\Tilde{\alpha}}$, whose value for a certain realisation $\boldsymbol{\Lambda}^*$ can for each \textbf{single} $\mathbf{x}^* \in S$ be interpreted as the probability that $\boldsymbol{\Lambda}^*$ is the "true" $\boldsymbol{\Lambda}$, i.e. the one for which it holds $f(\mathbf{x}^*,\boldsymbol{\Lambda}^*)=g(\mathbf{x}^*)$. As 
for any $\mathbf{x} \in S$, the "true" realisation can however not be known, especially if $\mathbf{x} \notin D \cap S$, the PDF $\pi_{f(\mathbf{x},\boldsymbol{\Tilde{\alpha}},\boldsymbol{\xi})}(\mathbf{\cdot})$ describes the probability on the "true" value searched for. 

The term \textit{approximately} here comes from the fact that there is no infinite data set $D$ used for the calibration, no infinite set of samples on $\boldsymbol{\Tilde{\alpha}}$ whose likelihood can be calculated, and that only PDFs are sampled that fit the way the parametrization of the embedded model error has been done, e.g. limited-order PCEs as introduced in Section~\ref{PCE}. Especially if no data error is considered, it is not an approximation of the full likelihood, which usually equals zero. Even, for the full likelihood, the model error concept is not useful, as the model error is zero while the likelihood is not and vice versa. 

As mentioned above, Sargsyan and coworkers recommend the use of moment-based likelihoods, which are computationally more tractable than sampling of the full or marginal PDFs \cite{Sargsyan2015,Sargsyan2019}. One of them, Approximate Bayesian Computation, is employed in this work and presented in the following. 

\paragraph{Approximate Bayesian Computation}
\phantom{(lala)}\\

In Approximate Bayesian Computation (ABC), the likelihood is approximated by a direct measure for comparing the statistics, usually the mean $\mu$ and standard deviation $\sigma$, of the model response to the actual data set $D$. 

This means we obtain a simultaneous minimization problem, where on the one hand, it is required for each $\left(\mathbf{x}^{(i)},y^{(i)}\right) \in D$ that 

\begin{equation}
    \mu^i_f(\bm{\tilde{\alpha}}) \approx y^{(i)},\\
\end{equation}
and on the other hand the remaining residual shall fit the standard deviation of the predition for each data point, 

\begin{align}
    \sigma^i_f(\bm{\tilde{\alpha}}) \approx |y^{(i)} - \mu^i_f(\bm{\tilde{\alpha}})|. 
\end{align}

To this end, a distance function $\bm{\rho}$ can be defined as
\begin{equation}
    \bm{\rho}(\bm{\tilde{\alpha}}) = \left(|y^{(1)} - \mu^1_f(\bm{\tilde{\alpha}})|,||y^{(1)}-\mu^1_f(\bm{\tilde{\alpha}})| - \sigma_f^1(\bm{\tilde{\alpha}})|,...,|y^{(L)} - \mu^L_f(\bm{\tilde{\alpha}})|,||y^{(L)}-\mu_f^L(\bm{\tilde{\alpha}})| - \sigma_f ^L(\bm{\tilde{\alpha}})|\right),
\end{equation}

which is weighted by a kernel function. 
Often, a Gaussian type kernel is used \cite{Sargsyan2015}:

\begin{equation}
K(\bm{\rho}) := \frac{1}{\sqrt{2\pi}} \exp\left(-\frac{||\bm{\rho}||_2^2}{2}\right). 
\end{equation}

The likelihood then is defined as \cite{Sargsyan2015}

\begin{equation}
    L(\bm{\rho}) = \frac{1}{\eta} K\left(\frac{\bm{\rho}}{\eta}\right),
\end{equation}

where $\eta$ is a tolerance parameter that determines how strict the moments have to match. 

In total, this yields the following ABC-likelihood:
\begin{equation}
L_{ABC}(\bm{\tilde{\alpha}}) = \frac{1}{\eta \sqrt{2 \pi}}\scalebox{1.2}{$\prod_{i = 1}^L$} \exp\left(\frac{(y^{(i)}-\mu_i(\bm{\tilde{\alpha}}))^2+(|y^{(i)}-\mu_f^i(\bm{\tilde{\alpha}})| - \sigma_f^i(\bm{\tilde{\alpha}}))^2}{2\eta^2}\right).
\label{LikABC}
\end{equation}

\subsubsection{Markov Chain Monte Carlo}
Markov Chain Monte Carlo (MCMC) methods are widely used for sampling from probability distributions, for which straight forward sampling methods are difficult. Especially in the context of Bayesian inference, MCMC is used to sample from the desired posterior distribution $\mathcal{P}(\bm{\tilde{\alpha}} \vert D)$. \\
MCMC combines Monte Carlo sampling with the construction of a Markov chain such that the stationary distribution of the Markov chain is the distribution from which the samples are desired. \\
The general MCMC procedure is summarized in the Metropolis-Hastings algorithm \cite{metropolis1953,hastings1970}. It relies on choosing new states based on some proposal distribution and accepting or rejecting the proposed states based on an acceptance ratio. \\
Within this acceptance ratio, the desired distribution has to be evaluated at the current state and at the proposed state. However, due to the ratio construction it suffices to evaluate a distribution proportional to the desired distribution, since the proportionality factor cancels in the ratio. Thus, for sampling the posterior distribution in the setting of Bayesian inference, it suffices to evaluate the product of likelihood and prior as $\mathcal{P}(\bm{\tilde{\alpha}} \vert D) \propto \mathcal{P}(D\vert \bm{\tilde{\alpha})} \mathcal{P}(\bm{\tilde{\alpha}}).$ \\
Extensions of the initial Metropolis-Hastings algorithm aim at improving the convergence behaviour of the algorithm. This is for example done by adaptively updating the proposal distribution, which results in an adaptive MCMC algorithm, rather than keeping it fixed throughout the entire process (\cite{haario2001}, \cite{haario2006}). 

\subsubsection{Predictive Moment Estimation \label{TB_PredMomEstim}}
The ABC likelihood construction requires only the computation of the predictive means $\mu_f^i(\bm{\tilde{\alpha}})$ and standard deviations $\sigma_f^i(\bm{\tilde{\alpha}})$, whereas the full PDF is not necessary. Using PCEs of the model outputs $f(\mathbf{x}^{(i)}, \bm{\Lambda}(\bm{\tilde{\alpha}},\bm{\xi}))$, this information can be easily obtained as follows \cite{Sargsyan2015,Sargsyan2019}: 
A PCE is constructed for each input $\mathbf{x}^{(i)}$ according to 
\begin{align}
    f(\mathbf{x}^{(i)}, \bm{\Lambda}(\bm{\tilde{\alpha}},\bm{\xi})) \approx \sum_{\ell=1}^P f_\ell(\mathbf{x}^{(i)}, \bm{\tilde{\alpha}})\Psi_\ell(\bm{\xi}),
\end{align}
where the coefficients $f_\ell(\mathbf{x}^{(i)},\bm{\tilde{\alpha}})$ are computed via integration by quadrature as 
\begin{align}
    f_\ell(\mathbf{x}^{(i)},\bm{\tilde{\alpha}}) &= \frac{1}{\Vert \Psi_\ell \Vert^2} \int_{\bm{\xi}} f(\mathbf{x}^{(i)},\bm{\lambda} +\bm{\delta}(\bm{\alpha}, \bm{\xi}^{(q)}))\Psi_\ell(\bm{\xi})\pi_{\bm{\xi}}(\bm{\xi})d\bm{\xi} \\
    &\approx \frac{1}{\Vert \Psi_\ell \Vert^2} \sum_{q=1}^Q w_q f(\mathbf{x}^{(i)}, \bm{\lambda} +\bm{\delta}(\bm{\alpha}, \bm{\xi}^{(q)}))\Psi_\ell(\bm{\xi}^{(q)})
\end{align}
with point-weight pairs $(\bm{\xi}^{(q)}, w_q)$ for $q=1,\dots,Q$.
The mean and the standard deviation can then be extracted as 
\begin{align}
    \mu_f^i(\bm{\tilde{\alpha}}) &= \mu_f(\mathbf{x}^{(i)},\bm{\tilde{\alpha}}) \approx f_0(\mathbf{x}^{(i)},\bm{\tilde{\alpha}}), \\
    \sigma_f^i(\bm{\tilde{\alpha}})^2 &= \sigma_f(\mathbf{x}^{(i)}, \bm{\tilde{\alpha}})^2 \approx \sum_{\ell=1}^P f_\ell(\mathbf{x}^{(i)}, \bm{\tilde{\alpha}})^2\Vert \Psi_\ell\Vert^2.
\end{align}

\subsection{Surrogate model construction \label{Surrogate}}
With the goal of alleviating the computational cost during the MCMC process, the model $f(\mathbf{x}, \bm{\lambda})$ can be approximated by surrogate models that are easier to evaluate than $f$ itself. To this end, for every operating condition $\mathbf{x}^{(i)}$, a surrogate model $f_s(\mathbf{x}^{(i)} , \bm{\lambda})$ given by a polynomial approximation is constructed according to 
\begin{align}\label{eq:surrogates}
    f(\mathbf{x}^{(i)}, \bm{\lambda}) \approx f_s(\mathbf{x}^{(i)},\bm{\lambda}) = \sum_{k=1}^K c_{ik}L_k(\bm{\lambda})
\end{align}
with multivariate Legendre polynomials $L_k(\bm{\lambda})$ \cite{Sargsyan2015,Sargsyan2019}. The coefficients $c_{ik}$ are then obtained via least-squares regression using samples $\bm{\lambda}^{(r)}$ within the parameter domain of $ \bm{\lambda}$ and the corresponding model evaluations $f(\mathbf{x}^{(i)}, \bm{\lambda}^{(r)})$. The obtained surrogates can then be used during the MCMC procedure for the moment prediction based on integration by quadrature in order to reduce the computational effort in each iteration.

\section{Methods}\label{Methodology}

\subsection{Chemical Reactor Network Modeling \label{Method_CRN}}
The numerical solution of a CRN consists in solving a large system of ordinary differential equations. The NetSMOKE framework presented in \cite{Trespi2021} is a CRN solver and has been used for obtaining CRN solutions within this work. It is based on the OpenSMOKE++ libraries \cite{CUOCI2015237}. 

\subsection{Uncertainty Quantification Framework \label{Method_UQframework}}
Within this work, the uncertainty quantification (UQ) toolkit UQTk (\cite{Debusschere2004}, \cite{debusschere2017}) is used, which comprises various UQ applications and libraries among others applications for surrogate construction, Bayesian inference employing model error embedding and sensitivity analysis. Thus, it is well suited for the purpose of this work and provides the main functionalities of the calibration procedure as defined in Table \ref{Cali_Prod_Overview}. \\
UQTk also contains readily compiled applications, which enables the easy use in combination with other software, see Section~\ref{Method_CombinedWorkflow}. 
Specifically, the applications \texttt{gen\_mi}, \texttt{regression} and \texttt{model\_inf} are employed. 
The utility \texttt{gen\_mi} creates a set of multi-indices that is used for the PC representation of the model surrogates that are constructed for each data point $\left(\mathbf{x}^{(i)},y^{(i)}\right) \in D$ as per \eqref{eq:surrogates}. The \texttt{regression} tool then computes the PC coefficients of the surrogates, which in return are provided as input to the \texttt{model\_inf} application that performs the Bayesian inference with embedded model error. Details on the usage and scope of the UQTk applications can be found in the UQTk manual \cite{UQTkManual}. In the following section, the coupling to the CRN solver framework NetSMOKE is described. 

\subsection{Combined CRN and UQ workflow \label{Method_CombinedWorkflow}}

The coupling of the CRN software NetSMOKE and the uncertainty quantification toolkit UQTk is implemented as a sequence of calls of compiled applications. These are either 
NetSMOKE or certain UQTk applications, with the respective inputs required by the programs and adapted to the given task. This means the coupling has been done externally without having to change the source code of NetSMOKE or UQTk. Only shell and python \cite{PythonRef} scripts have been used for calling the programs as appropriate or doing the data analysis. 

The CRN software NetSMOKE has been used to provide the training data for the surrogate construction (see Section~\ref{Surrogate}) in the first step. After the calibration procedure, which yields the posterior distributions for the free CRN model parameters, it has been used to evaluate CRNs for a given set of operation conditions 
and each 
posterior sample of the model parameters. This yields the posteriors of the CRN prediction. 

After the surrogate training data has been constructed with the help of NetSMOKE, this data is transferred to UQTk and the calibration steps described in Section~\ref{TB_CaliP}, for the theory part, and Section~\ref{Method_UQframework} are applied. In Table~\ref{Method_DefaultCaliOptions}, the default calibration options which can be specified in the used applications of UQTk, are given. These have been used for all the calibrations presented in this work. Some other parameters depend on the given calibration task. These will be specified along the detailed description of the calibration problems in Section~\ref{Results}. 

\begin{table}[h!]
    \centering
    \begin{threeparttable}
        
    \begin{tabular}{ll}
    \hline
    Surrogate polynomial order & 6$^{*}$ \\
      Number of MCMC samples   & 400 000$^{**}$ \\
      Burn-in (see \cite{UQTkManual})  & 200 000\\
      Thinning factor (see \cite{UQTkManual}) & $\frac{1}{10}$\\
      Length of thinned MCMC chain& 20002 \\
      Prior type & Uniform\\
      PCE type & Legendre-uniform, UQTk option\\
      & "full" pdf type ($\equiv$Eq.~\eqref{TB_TruncatedPCE})\\
      Number of $\bm{\xi}$-samples, for evaluating the PDF of $\bm{\Lambda}(\bm{\Tilde{\alpha}},\bm{\xi})$& 100\\
      Number of $\bm{\Lambda}$-samples used to evaluate CRN prediction in NetSMOKE,&10 000\\
      to obtain full posterior PDF evaluation& \\
      \hline
    \end{tabular}
    \begin{tablenotes}
    \item $^{*}$ The quality of the PCE fits has been confirmed for each PCE constructed. 
        \item $^{**}$ Each MCMC run has been repeated once, using the MAP values of the first run as initial guess, to check for convergence.
    \end{tablenotes}
            \end{threeparttable}
    \caption{Default options used during calibration procedures. These are used for any case where it is not stated otherwise. }
    \label{Method_DefaultCaliOptions}

\end{table}

The UQTk application \texttt{model\_inf} is the last one in the chain of UQTk applications used. It yields many different outputs, among them the mean 
of the calibrated parameters and the model predictions for all the data points, for which PCE surrogates and operation conditions have been provided. This especially comprises the operation points actually used for the calibration, but additional data points can be specified, too. 
The mean of the calibrated model parameters are given by \cite{Sargsyan2015,Sargsyan2019}

\begin{equation}
    \mu_{\text{parameter}} = \left[ \left[ \bm{\Lambda}(\bm{\Tilde{\alpha}},\bm{\xi} )\right]_{\bm{\xi}}\right]_{\bm{\Tilde{\alpha}}},
\end{equation}
where $\left[ x \right]_y$ denotes the expectation of $x$ with respect to $y$. Accordingly, for the model prediction, which also depends on the operating condition $\mathbf{x}$, the mean value is given by

\begin{equation}
        \mu_{\text{prediction}} = \left[ \left[ f(\mathbf{x},\bm{\Lambda}(\bm{\Tilde{\alpha}},\bm{\xi} ))\right]_{\bm{\xi}}\right]_{\bm{\Tilde{\alpha}}}. 
\end{equation}

The program \texttt{model\_inf} also yields and the posterior mean value and the posterior MAP value of the variances of the random variables $\bm{\Lambda}({\bm{\Tilde{\alpha}}})$ and the posterior variance of the mean of $\bm{\Lambda}({\bm{\Tilde{\alpha}}})$ 
and the same for the predictions for those operating conditions given to UQTk. The variances are given by

\begin{align}
        \sigma_{\textrm{parameter, mean}} = \left[ \text{Var}\left( \bm{\Lambda}(\bm{\Tilde{\alpha}},\bm{\xi} )\right)_{\bm{\xi}}\right]_{\bm{\Tilde{\alpha}}}&(\text{posterior mean of variance}),\\
        \sigma_{\textrm{parameter, variance}} = \text{Var}\left( \left[\bm{\Lambda}(\bm{\Tilde{\alpha}},\bm{\xi} )\right]_{\bm{\xi}}\right)_{\bm{\Tilde{\alpha}}}&(\text{posterior variance of mean}),
\end{align}
where $\text{Var}\left(x\right)_y$ denotes the variance of $x$ with respect to $y$. For the predictions, it holds 

\begin{align}
        \sigma_{\textrm{prediction, mean}} = \left[ \text{Var}\left( f(\mathbf{x},\bm{\Lambda}(\bm{\Tilde{\alpha}},\bm{\xi} ))\right)_{\bm{\xi}}\right]_{\bm{\Tilde{\alpha}}}&(\text{posterior mean of variance}),\\
        \sigma_{\textrm{prediction, variance}} = \text{Var}\left( \left[f(\mathbf{x},\bm{\Lambda}(\bm{\Tilde{\alpha}},\bm{\xi} ))\right]_{\bm{\xi}}\right)_{\bm{\Tilde{\alpha}}}&(\text{posterior variance of mean}). 
\end{align}

Also, the program yields the full MCMC chain and the thinned MCMC chain, which only starts after the burn-in samples and is thinned by a thinning factor, see Table~\ref{Method_DefaultCaliOptions}. 
More details on the outputs can be found in the UQTk manual \cite{UQTkManual}. 

For the calibration results presented in Section~\ref{Results}, the mean predictions and standard deviations have directly been taken or calculated, respectively, from the UQTk output. The standard deviation $\sigma$ has been calculated by 

\begin{equation}
    \sigma = \sqrt{\sigma_{\textrm{mean}}^2+\sigma_{\textrm{posterior}}^2}, 
\end{equation}

where $\sigma_{\textrm{mean}}^2$ is the posterior mean of the variance and $\sigma_{\textrm{posterior}}^2$ is the posterior variance of the mean. Both, $\sigma_{\textrm{mean}}^2$ and $\sigma_{\textrm{posterior}}^2$ can be with respect to the model parameters or the predictions as introduced above. For the variance additivity it is referred to \cite{Sargsyan2015}. 

The parameter posterior visualization is conducted with help of the thinned MCMC chain, which contains the (thinned) MCMC samples for all model parameters stored in the vector $\bm{\Tilde{\alpha}}$ (see Section~\ref{TB_Model}). For every $\bm{\Tilde{\alpha}}$-sample, the random variable $\bm{\Lambda}$ parameterized by $\bm{\Tilde{\alpha}}$ is sampled according to its PCE representation, and the range and distribution of $\bm{\xi}$. This is done with the sampling routine of the python package \texttt{numpy} \cite{NumpyHarris2020}. On these samples, kernel density estimation (KDE) is applied in order to obtain a PDF on the parameters. This KDE is obtained with the kde routine of the python \texttt{scipy} package \cite{2020SciPy-NMeth} and used for the PDF plots in Section~\ref{Results}. 

For computing the predictions from the posterior distributions of the model parameters, the samples are uniformly thinned out again, regarding the large number of samples, which is the product of the thinned MCMC number of steps with the number of samples per $\bm{\Lambda}$ parameterized by $\bm{\Tilde{\alpha}}$. The thinned parameter samples are then provided to NetSMOKE to calculate the prediction samples. Again KDEs are computed for plotting and also to enables numeric integration of the PDF function.

\section{Flash Ironmaking \label{Testcase}}

Flash ironmaking is a novel technology for the direct reduction of iron ore concentrate (mainly magnetite (\ce{Fe3O4})) to elemental iron (\ce{Fe}) with enormous potential to drastically reduce \ce{CO2} emissions associated with the steel industry \cite{Sohn2023}. This is achieved by using hydrogen (\ce{H2}) or hydrogen-natural gas (mainly methane (\ce{CH4})) mixtures as reducing agents, thus avoiding a coke-based reduction process in blast furnaces. The main controlling reactions in the case of hydrogen are \cite{Sohn2023}:
\begin{align}
    \ce{Fe3O4 (s) + 4H2 (g) &<-> 3Fe (s) + 4H2O (g)} \\
    \ce{FeO (s) + H2 (g) &<-> Fe (s) + H2O (g).}  
\end{align}
Due to the endothermic nature of magnetite reduction with hydrogen, sufficient heat must be provided to sustain the reaction. This is achieved by partial oxidation (combustion) of the reducing agent, in this case hydrogen and/or methane.
Magnetite is available in powder form with particle sizes less than 100 µm. The small particle size results in high reaction rates, allowing almost complete reduction within a few seconds of residence time in the reduction reactors \cite{Sohn2023}. 

The reduction of magnetite under process conditions relevant to the flash ironmaking process is investigated by Sohn \textit{et al.} \cite{Sohn2023,Sohn2021,Sohn2021a,Abdelghany2019,Abdelghany2020,Fan2016,Chen2015} in a series of studies at various scales. First, a kinetic expression is obtained from experiments in a laminar flow reactor. In order to realize a scale-up, intermediate scale tests were carried out in a laboratory flash reactor. Finally, a smaller pilot reactor was operated and CFD simulations for a pilot reactor close to industrial conditions has been conducted by Sohn and co-workers.

\begin{figure}[]
\centering
\includegraphics[width=0.55\textwidth]{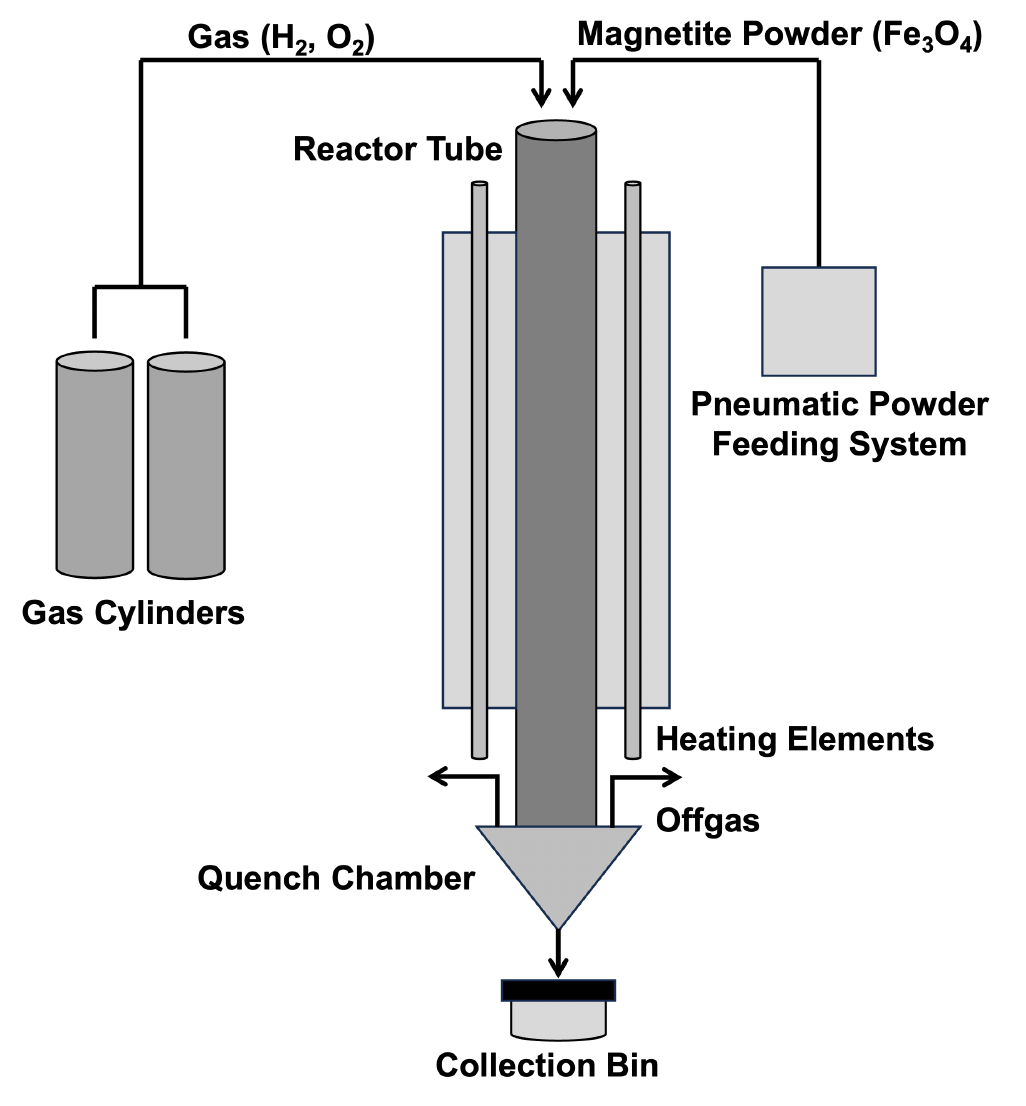}
\caption{Simplified schematic of the laboratory flash reactor for magnetite reduction, adapted from \cite{Sohn2023}.} 
\label{fig:FlashReactor}
\end{figure}

The CRN study presented in this work focuses on the laboratory scale flash reactor. A schematic of the apparatus is shown in Fig.~\ref{fig:FlashReactor}. The magnetite powder is fed into the 213~cm long reactor tube of 19.5~cm width by a pneumatic feed system along with the gaseous fuel/oxidizer mixture. Electrical heating is provided by heating elements placed in a furnace around the reactor. Reduced samples are collected in a collection bin after quenching the hot flue gas to reduce the amount of particle loss in the flue gas. The degree of reduction is then determined by inductively coupled plasma mass spectroscopy. During the experiments, the particle residence times, the amount of remaining reducing gases and the flame configuration were varied and their effect on the degree of reduction was studied. 

In addition to the experimental studies, a three-dimensional CFD simulation of the laboratory flash reactor has been performed. The gas phase is modeled in an Eulerian framework, while a Lagrangian description is used to describe the particle phase within a Reynolds Averaged Navier Stokes (RANS) simulation. The simulation results are validated with the experiments and provide details on the flow and temperature fields inside the reactor. More information on the experimental and numerical studies is available in \cite{Sohn2023}.

In the present study, the results obtained for the experiments with hydrogen as the only reducing agent have been used for calibration of a CRN model. Sohn and co-workers have found hydrogen to be the driving species for reduction even if carbon-monoxide is present, and testing the calibration is easier with a simpler configuration, too. Out of the various experimental operating conditions reported by Sohn, those with the same flame configuration (HOH, see \cite{Sohn2023}) and powder-feeding mode (side slots) have been chosen for the present study to reduce the number of variable parameters in a first modeling attempt. 
Table~\ref{TC_OPTable} reports the chosen selection of operating points with respect to gas and particle mass flows along the respective measured reduction degrees.

\begin{table}[]
    \centering
    \begin{tabular}{ccccc}
     Operation Point & $\text{H}_2$ flow (l/min) & $\text{O}_2$ flow (l/min) & $\text{Fe}_3\text{O}_4$ flow (g/min) & Reduction Degree (\%) \\
     \hline
     \hline
     A & 15.3 & 2.16 & 1.9 & 82 \\
     B & 15.3 & 2.36 & 1.7 & 76 \\
     C & 15.3 & 2.48 & 2.0 & 70 \\
     D & 15.3 & 2.50 & 1.9 & 70 \\
     E & 15.3 & 2.72 & 2.1 & 57 \\
     F & 20.0 & 2.20 & 2.2 & 96 \\
     G & 20.0 & 2.96 & 1.8 & 84 \\ 
     H & 20.0 & 2.96 & 2.2 & 80 \\ 
     I & 20.0 & 3.22 & 1.9 & 80 \\
     J & 20.0 & 3.70 & 2.0 & 63 \\ 
     K & 40.0 & 4.32 & 2.2 & 92 \\ 
     L & 40.0 & 6.40 & 2.1 & 77 \\ 
     M & 60.0 & 9.65 & 1.9 & 74 \\ 
     N & 60.0 & 9.65 & 2.2 & 74 \\ 
     O & 60.0 & 10.65 & 2.3 & 72 \\ 
     P & 60.0 & 11.20 & 2.0 & 64 \\ 
     Q & 60.0 & 12.20 & 2.0 & 49 \\ 
     \hline
     R& 20.0 & 3.50 & 2.0 & - \\
     \hline
     \hline
    \end{tabular}
    \caption{Operation points selected from the experimental studies by Sohn \cite{Sohn2023} with the relevant operating conditions and experimentally obtained reduction degrees. OP R is not from the Sohn study such that there is no validation data available from experiment or CFD, and has been introduced for testing consistency between the modeling regimes that will be introduced in Section~\ref{RS_ModelingTFlame}. The particles always are fed into the reactor by a $\text{N}_2$ flow of \SI{2.8}{\liter\per\minute}.} 
    \label{TC_OPTable}
\end{table}

\subsection{Kinetic mechanism}

As mentioned before, Sohn \textit{et al.} have developed kinetic mechanisms for the reduction of hematite and magnetite powders in different reducing atmospheres \cite{Chen2015,Fan2016a,Fan2017,Sohn2023}. 

For magnetite reduction with temperatures up to \SI{1623}{\kelvin}, a global rate expression has been obtained for the reduction degree, which is a dimensionless measure for the progress of the reduction given by \cite{Sohn2023}: 

\begin{equation}
    X = \frac{m_\textrm{i}-m_\textrm{t}}{m_\textrm{i}Y_{\textrm{O,sol,i}}},
    \label{TC_DefinitionRedDeg}
\end{equation}

where $X$ is the reduction degree defined on $\left[0,1\right]$, $m_\textrm{i}$ is the initial solid mass of the experimental sample (not to be confused with a statistical sample in the present calibration context) and $m_\textrm{t}$ the sample mass after reduction and $Y_{\textrm{O,sol,i}}$ is the mass fraction of oxide bounded in the sample before reduction, where the mass fraction is with respect to the total solid sample mass, not to be confused with the general mass fractions stated in the CRN context, e.g. Equations~\eqref{PSR_massfrac_equation} and~\eqref{PFR_massfrac_equation}. 

The global rate equation for the reduction degree is given by the following initial value problem \cite{Sohn2023}:~
\begin{align}
    \label{TC_ReductionDegreeRateEquation}
    \frac{dX}{dt} = a \cdot\exp\left[ \frac{-E_\textrm{A}}{R T}\right]\cdot\left[ p_{\textrm{H}_2}-\frac{p_{\textrm{H}_2\textrm{O}}}{K}\right]\cdot(1-X), \\
    X(0) = X_0, 
\end{align}

where $p_{\textrm{H}_2}$ and $p_{\textrm{H}_2\textrm{O}}$ are the hydrogen and steam partial pressures, measured in atm, $K$ is the equilibrium constant, $a$ is an Arrhenius-type pre-factor, and  $E_\textrm{A}$ is an activation energy. The experimentally determined values provided by Sohn are $a = 1.23\cdot10^7\frac{1}{\textrm{atm}}$ and $E_\textrm{A} = \SI{196000}{\joule\per\mole}$. It holds $X_0 = 0$ throughout this work. The equilibrium constant depends on the temperature and has experimentally been determined by Fan \cite{Fan2016a} for five different temperatures. In order to obtain the equilibrium constant outside the five temperatures tested, a linear fit has been conducted, using the proportionality relation $\log(K) \sim \frac{1}{T}$ \cite{Fradet2023}, which has shown to predict the experimental values very well. 

The global kinetic equation~\eqref{TC_ReductionDegreeRateEquation} gives no information about the fractions of different iron or iron oxide phases during reduction, and according to \cite{Chen2015} these are in general not known during experiments. For the numerical CRN solution, Equation~\eqref{TC_ReductionDegreeRateEquation} are transformed to a set of species equations, including magnetite and pure iron, however, without changing the global behavior for the reduction degree and gas species as given in \eqref{TC_ReductionDegreeRateEquation}. 
Equation~\eqref{TC_DefinitionRedDeg} has accordingly been transformed to 
\begin{equation}
    X = 1-\frac{Y_{\text{Fe}_3\text{O}_4}}{Y_{\text{Fe}_3\text{O}_4,{\text{in}}}},
    \label{TC_RedDegNSpp}
\end{equation}
where $Y_{\text{Fe}_3\text{O}_4}$ and $Y_{\text{Fe}_3\text{O}_4,{\text{in}}}$ are the current and initial mass fractions of magnetite, now with the mass fractions being relative to the total mass, including gas species. These are the mass fractions that are obtained from the CRN solver. 
The reduction degree and potentially the gas species, are the only quantity of interest in the given case, and Equation~\eqref{TC_RedDegNSpp} has been used to obtain the reduction degree from the mass fractions computed with the CRN solver. 

Furthermore, the partial oxidation of hydrogen has not been explicitly modeled in the present study; it has been assumed that all oxygen is consumed so fast that this does not have any (chemical) influence on the reduction process. This has been motivated by CFD results for the species distribution in the reactor presented in \cite{Sohn2023}. Contrarily, the influence of the flame on the temperature distribution in the reactor has an important role and is investigated in Section~\ref{Results}.

\subsection{Design space \label{TC_DesSpace}}
\begin{figure}[h!]
    \centering
    \includegraphics[width=0.7\textwidth]{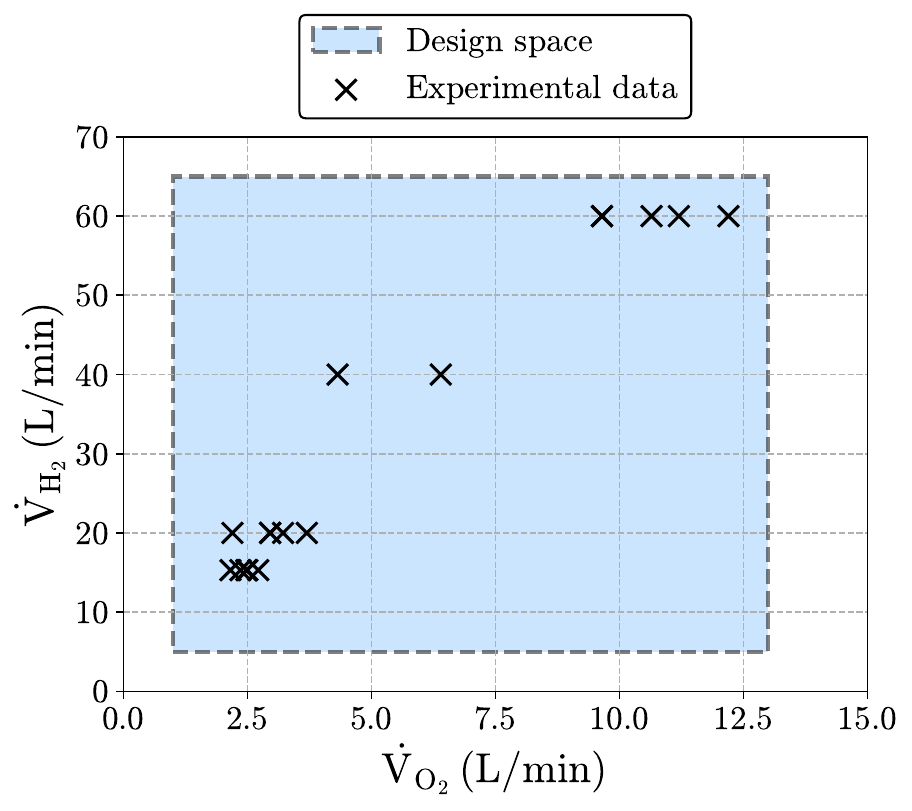}
    \caption{Exemplified subset of the design space of operating conditions or the given test case. A valid CRN model shall be calibrated on the full design space. However, for the present case, the exact definition of the design space is difficult, for details it is referred to the explanation in Section~\ref{TC_DesSpace}. }
    \label{TC_DesignSpace}
\end{figure}

The design space defines the set of operating conditions on which a valid CRN model shall be calibrated. For the laboratory reactor considered in the present work, the relevant operating conditions are the mass flow of magnetite provided to the reactor as well as the hydrogen and oxygen volumetric flows. Moreover, for the present case it is assumed that the hydrogen and oxygen flows have the biggest influence on the CRN model itself, which will be extensively discussed in Section~\ref{Results}. In Figure~\ref{TC_DesignSpace}, a projection of the full design space onto the $\Dot{V}_{\text{H}_2}$-$\Dot{V}_{\text{O}_2}$-plane is exemplified. However, the exact bounds of the design space are difficult to define for the present case, as there is a correlation between the hydrogen and the oxygen flow for those operation points belonging to the experimental data set. This correlation can be understood since both, too elevated and too low oxygen flows compared to the hydrogen flow would be problematic. If the ratio of oxygen flow to hydrogen flow is too high, there will be no hydrogen left for the actual reduction of magnetite after the partial combustion. If it is too low, the flame will probably not stabilize. The exact definition of the design space for an application case where the available data is non-uniformly spread will be revisited in future work. The focus of the present work is on the actual development of CRN models being valid for, at least, many operation points at once. 

\section{Results}\label{Results}

In the following section, the results of the present study are presented and discussed. This includes the construction of a CRN model to describe the laboratory reactor by Sohn \cite{Sohn2023} for various operating conditions, preparatory studies for enhancing the model and finally the calibration of the identified sensitive model parameters. The role of the different types of uncertainty in the present case will also be discussed. An overall discussion on the results and the scope of this study in relation to the objectives of this work follows afterwards in Section~\ref{Discussion}. 

\subsection{Set up of a CRN model for the lab reactor \label{SetUpCRN}}

\begin{figure}
    \centering
    \includegraphics[width = 0.99\columnwidth]{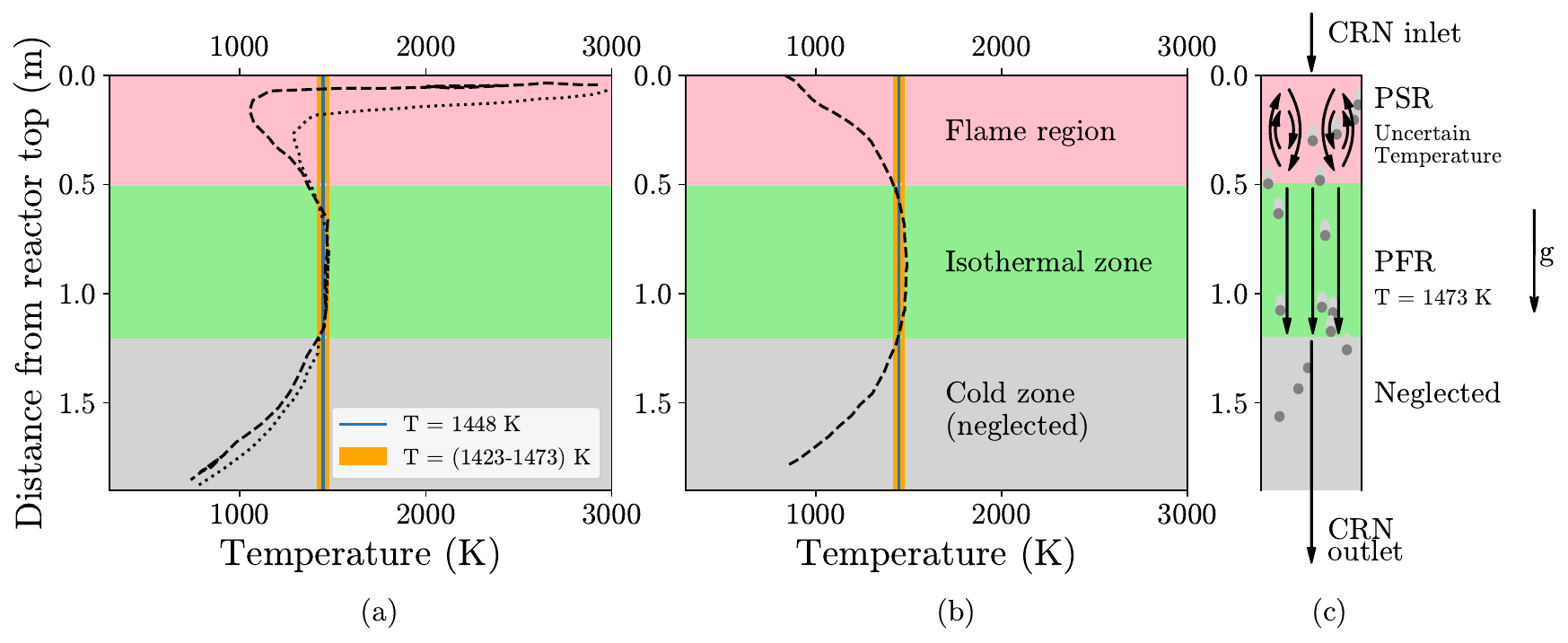}
    \caption{(a) Approximate temperature profile at the reactor center line obtained in the CFD simulations by Sohn \textit{et al.}, for different operating conditions, adapted from \cite{Sohn2023}, (b) Approximate temperature profile at the reactor wall obtained in the experiments by Sohn \textit{et al.}, adapted from \cite{Sohn2023}, (c) Simple CRN for modeling the lab reactor, consisting of a perfectly stirred reactor (PSR) for modeling the turbulent flame zone and a plug flow reactor (PFR) for modeling the laminar isothermal zone of the reactor. The particle slip due to gravity leads a lower particle residence times compared to the gas phase. }
    \label{RS_CRN_SohnLab}
\end{figure}

In the present section, a CRN model is set up, which models 
the lab reactor by Sohn \textit{et al.} \cite{Sohn2023} presented in Section~\ref{Testcase}. 
The secondary data available in the literature \cite{Sohn2023,Sohn2021,Sohn2021a,Abdelghany2019,Abdelghany2020,Fan2016}, which consists of experimental and CFD data, is used for the development of a CRN model. Hereby, the temperature distribution, flow structure and particle slip are of particular importance in order to account for the reaction kinetics and residence time. 
In Figures~\ref{RS_CRN_SohnLab}(a) and~\ref{RS_CRN_SohnLab}(b), the simulated temperature profile at the reactor center line, and the measured temperature profile at the reactor wall for different operating conditions, as presented in \cite{Sohn2023}, are shown. Sohn and co-workers describe an isothermal temperature region of approximately \SI{70}{\centi\meter} length in the middle of the reactor, seen in axial direction, that is approximately constant for all operating conditions and throughout the radial direction of the reactor. This can also be seen in Figures~\ref{RS_CRN_SohnLab}(a) and~\ref{RS_CRN_SohnLab}(b) and according to Sohn the temperature is $1448 \pm$\SI{25}{\kelvin}. This zone is denoted as an isothermal reaction zone. It is located between approximately \SI{50}{\centi\meter} and \SI{120}{\centi\meter} away from the top line of the reactor. 

Above the isothermal reaction zone, the flame for partial oxidation of hydrogen is situated. Therefore, this region will be referred to as the flame region in the following. As can be seen in Figure~\ref{RS_CRN_SohnLab}(a), there is a huge temperature gradient at the center line, where the hottest part of the flame is located. The exact temperature profile, including the spatial extent of the flame, strongly depends on the operating condition. In total numbers, temperatures between approximately \SI{1100}{\kelvin} and \SI{3100}{\kelvin} are reached. %

At the reactor walls, a different behavior can be observed, as can be seen in Figure~\ref{RS_CRN_SohnLab}(b). There, an approximately linear increase of temperature, starting from \SI{800}{\kelvin}, and finally reaching the isothermal temperature of \SI{1448}{\kelvin}, is visible. From this data, it can be concluded that the exact spatial temperature in the flame region is strongly dependent on the operating condition and in general not known without access to primary data. In addition, Sohn reports a turbulent zone in the upper part of the reactor, while there is laminar flow in the lower part (see \cite{Sohn2023}, p. 164). 
Thus, the particles will disperse in the flame region, and it remains uncertain to what extent they pass through the hottest areas of the flame when introduced via the side slots 
(see Section~\ref{Testcase}). 

This is why in a first CRN modeling attempt, this region is modeled as one compartment, whose temperature is treated as a (highly) uncertain model parameter. The isothermal region is modeled as another compartment of the CRN. 
As there is a turbulent zone in the upper part of the reactor, 
the flame region is modeled by a perfectly stirred reactor (PSR) modeling block, whereas the isothermal zone is modeled by a plug flow reactor (PFR) modeling block as laminar flow is dominant further down the reactor. From the available data, it is not entirely clear if the transition from turbulent to laminar flow regimes exactly occurs at \SI{50}{\centi\meter} from the top of the reactor, where the boundary between flame and isothermal zone has been defined, but this assumption is at least not unreasonable (cf. \cite{Sohn2023}, p. 164). 

Finally, a strong decrease in temperature both at the reactor center line and near the walls is observable further down the reactor, below the isothermal region, in Figures~\ref{RS_CRN_SohnLab}(a) and~\ref{RS_CRN_SohnLab}(b), and Sohn also assumes that there is no significant reaction ongoing in that region \cite{Sohn2023}. Therefore, this region is neglected in the CRN model developed here. However, especially for the temperature profile measured near the reactor walls, it is visible that the temperature even exceeds \SI{1473}{\kelvin}, and that temperatures higher than \SI{1423}{\kelvin} are measured at locations even a bit lower than the boundary of the isothermal zone. For this reason, a temperature of \SI{1473}{\kelvin} is assumed for the isothermal zone. 

All together, 
a generic CRN model 
consisting of a PSR and a PFR in series, where the PSR temperature is not known a priori and will depend on the operating conditions, as will be seen later on, and the PFR temperature is \SI{1473}{\kelvin} 
is proposed. 

As the reactor has a vertical orientation (see Figure ~\ref{fig:FlashReactor}), Sohn reports a particle slip due to gravitational acceleration of the particles \cite{Sohn2023}. This leads to strongly reduced residence times for particles compared to the gas phase for many operation points and thus has to be taken into account. 
The particle terminal velocity $u_t$, which has to be added to the gas velocity, is given by \cite{Sohn2023}

\begin{equation}
    u_\tm{t} = \frac{d^2_\tm{p} g (\rho_\tm{p} -\rho_\tm{g}) }{18 \mu},
\end{equation}
where $d_\tm{p}$ is the particle diameter, g is the gravitational constant, $\rho_\tm{p}$ and $\rho_\tm{g}$ are the particle and gas density respectively and $\mu$ is the gas viscosity. In Sohn's experiments, the average particle diameter is \SI{32.5e-06}{\meter} and the initial particle density is \SI{5150}{\kilo\gram\per\cubic\meter} according to the report in \cite{Sohn2023}. 

As the objective of the current work is to construct a model that is usable for any operating condition in the design space, an ideal gas approach is used to calculate the gas viscosity. Correction factors occuring due to the mixture of different gas components have been taken into account according to \cite{Davidson1993,Mason1958},~
\begin{align}
    \mu_{i} = \sqrt{\frac{8 k_\textrm{B} T M_{i}}{9 \pi^3 \left(d_{\tm{g},i}\right)^4}},\\
    \mu_{\tm{gas}} = \sum_i\frac{ x_i \mu_i}{\sum_j x_j \Phi_{ij}},\\
    \Phi_{ij} = \frac{1}{2\sqrt{2}}\left(1+\frac{M_i}{M_j}\right)^{-\frac{1}{2}}\cdot \left(1+\left(\frac{\mu_i}{\mu_j}\right)^{\frac{1}{2}}\cdot \left(\frac{M_j}{M_i}\right)^{\frac{1}{4}}\right)^2.
\end{align}

In the above equations, $i$ and $j$ denote a single species, and "gas" denotes the mixture of gas species. The symbol $k_\tm{B}$ denotes the Boltzmann constant, $T$ is the temperature, $M$ is the atomic mass, $d_\tm{g}$ is the kinetic diameter of the gas species and $\Phi_{ij}$ is a correction factor for the mixture introduced by Mason and Saxena \cite{Mason1958}. 

This simplified assumption leads to slight deviations from the residence times calculated by Sohn, which has however not shown to be crucial for the overall modeling approach. Possible future enhancements of the given approach will be discussed in Section~\ref{RS_OP_F}. The reactor models in the CRN have been solved with respect to the calculated particle residence times. The differing gas and particle velocities have not explicitly been modeled, as the relative change in the gas composition due to the reduction reaction is only in the order of a few percent and it has been observed that the initial ratio of hydrogen to steam is the most relevant regarding the final reduction degree. 

The proposed CRN is schematically depicted in Figure~\ref{RS_CRN_SohnLab}(c). 
It has to be noted that the proposed CRN structure is very simple, even in the scope of CRN models being strongly reduced models compared to CFD or experiments. 
However, with the objective of this work being the calibration of a CRN on the full design space of operating conditions, the extent, to which details that are strongly dependent on the operating conditions can be modeled in a meaningful way, is limited. 
Moreover, through employing uncertainty quantification, it will be possible to evaluate the quality of this simple model afterwards. 

Before calibration of the CRN model, it is tested for single operation points, and results are reported in the following section. This serves as a partial validation of the model and a preparatory study for later calibration.

\subsection{Calibration of the PSR temperature for single operation points \label{RS_ClassicTCali}}
In order to evaluate in how far a CRN model could be calibrated on the full design space, a preparatory study has been conducted. This also serves as a partial validation of the CRN model constructed in Section~\ref{SetUpCRN}, as will be seen below. 

For every operation point considered here (see Table~\ref{TC_OPTable}), a classical calibration of the PSR temperature, which is the only undetermined parameter in the present CRN model, has been conducted. Classical here means that no model and no data error has been considered during the calibration. When only single operation points are used for the calibration, the model error concept would not work anyway and objective of the preparatory one-to-one calibration presented here only is to enable an estimate on how well the CRN model works and to examine how it could be globally calibrated. Note that throughout this work, the target value of the calibration is always the reduction degree obtained during the flash ironmaking process. 

\begin{figure}[h]
    \centering
    \includegraphics[width=0.7\columnwidth]{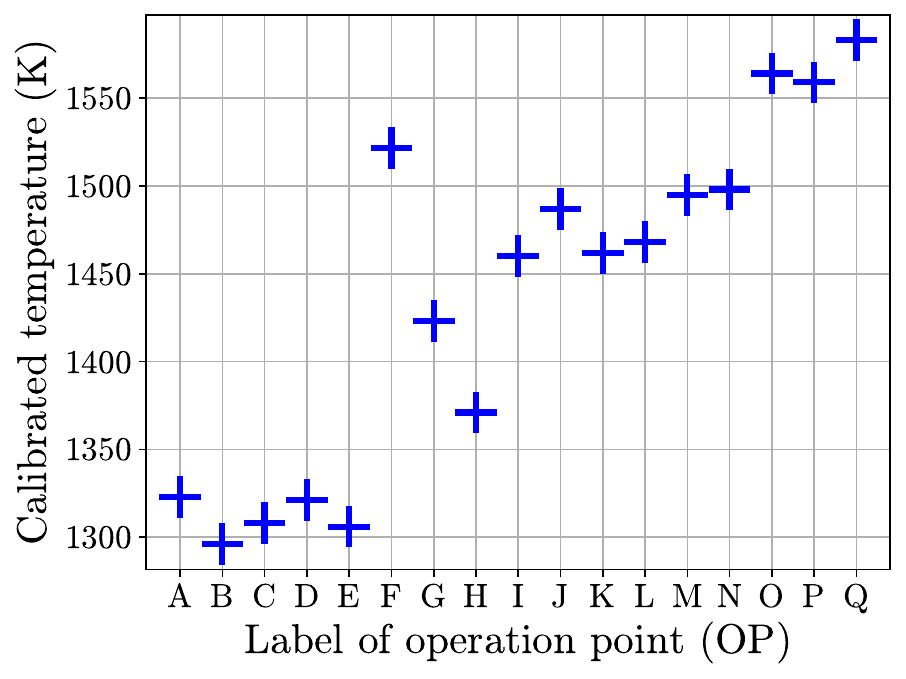}
    \caption{Calibrated temperatures for each single operation point (see Table~\ref{TC_OPTable}). Backgrounds and calibration details are given in the text. }
    \label{RS_CalibratedT_ListPlot}
\end{figure}

In Figure~\ref{RS_CalibratedT_ListPlot}, the results of the classical calibration for each operation point are shown. Note that there is no link to any physical variable in that plot, so no functional temperature model can be deduced from that representation. This will be achieved by connecting the data to the respective operating conditions in a later step. 

However, the following can be learned from Figure~\ref{RS_CalibratedT_ListPlot}. Firstly, the calibrated temperatures are in a range between \SI{1296}{\kelvin} and \SI{1583}{\kelvin} and are more or less evenly spread within that range. This leads to the direct conclusion that it will not be possible to calibrate a single PSR temperature on the full design space, and more parameters are needed. Secondly, it can be stated that the calibrated temperature values are consistent and in a reasonable order of magnitude. Overall, a global increase of the calibrated temperature, even though not monotonically, is observed for increasing total gas flow rate (see Table~\ref{TC_OPTable}). This is reasonable as a higher gas flow will lead to a larger flame, which is also reported by Sohn \cite{Sohn2023}. Next, it is observed that for some operating conditions the PSR temperature is lower as the temperature in the isothermal region, and for some it is higher. In total, the spread is $\sim\pm\SI{200}{\kelvin}$ around the isothermal temperature of $1448\pm\SI{25}{\kelvin}$ ($\SI{1473}{\kelvin}$ used as PFR temperature, see Section~\ref{SetUpCRN}). Although the exact temperature profiles within the PSR region are not known, the profiles reported in Figures~\ref{RS_CRN_SohnLab}(a) and~\ref{RS_CRN_SohnLab}(b) indicate that there are regions with both, strongly lower and higher temperatures present in that zone. 
The calibrated temperature can be seen as a kinetically averaged temperature similar as introduced in \cite{Faravelli2001}, which accounts for the nonlinear temperature dependence of the governing kinetics and does not have to be equal to the actual (volumetric) average gas or particle temperatures. 

The results have shown that there actually is a solution for the PSR temperature for every single operation point. This would not have been the case, if the particles would already reach the experimental value of the reduction degree when only travelling through the PFR zone, which would imply that every physically reasonable PSR temperature would lead to overshooting reduction degrees. On the one hand, this shows that the calibration problem is well-posed, and on the other hand this, together with the considerations above, can be seen as a partial validation of the CRN model, as there are no indicators for the model not being usable for the given task. 

\subsection{Modeling of temperature in the flame region \label{RS_ModelingTFlame}}

As has been seen in Section~\ref{RS_ClassicTCali}, it is not possible to calibrate a single PSR temperature for covering the full design space of operating conditions. This holds, even if a model error is considered, as the physical problem is known well enough to know that results mostly are sensitive to temperature, and a range of $\sim\SI{300}{\kelvin}$ will not lead to meaningful results in terms of reduction degrees. 

Therefore, more or different model parameters are needed for accurately modeling the reduction degrees that can be obtained in the lab reactor across various operation points. One possible approach would be to increase the number of CRN parameters, e.g. the number of modeling blocks, which is usually done in the CRN context, see e.g. \cite{Trespi2021}. However, as discussed above, the present problem is different, as it is not the CRN per se, which is not able to model the experimental data for given conditions, but the task is to find a model being valid for many operating conditions at once. This is why the authors propose to model the temperature itself, by the introduction of meta parameters to the CRN model. In order to achieve a suitable temperature model, it has to be investigated how the temperature depends on the operating conditions, that is, the species mass flows provided to the reactor. 

It is assumed that the hydrogen and oxygen flow rates are the parameters that have the strongest influence on the average  
temperature in the flame region. 

\begin{figure}[h]
    \centering
    \includegraphics[width=0.7\columnwidth]{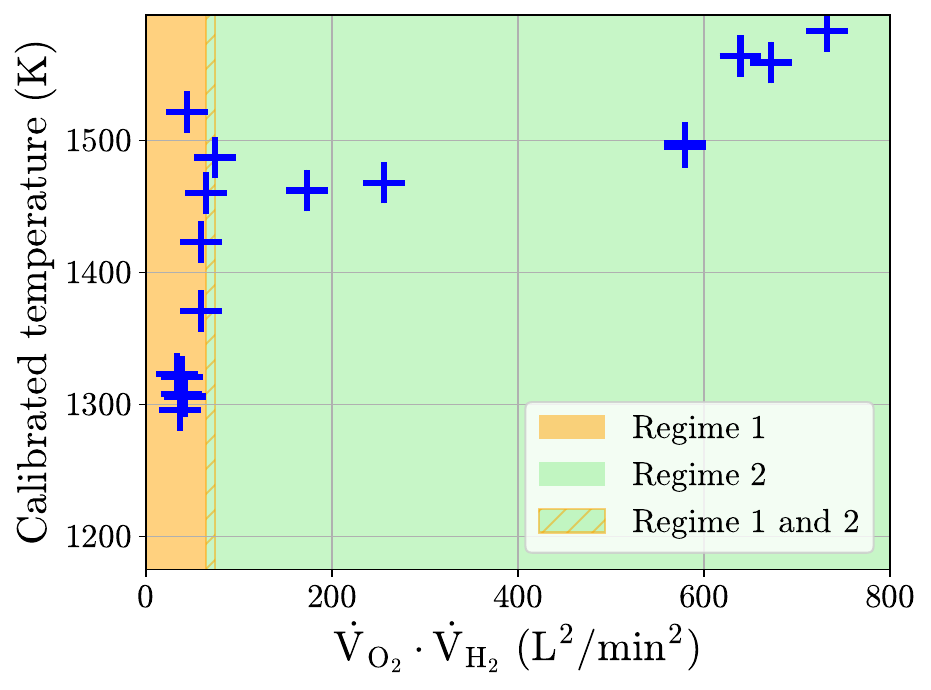}
    \caption{Temperatures calibrated for single operating conditions in Section~\ref{RS_ClassicTCali}, plotted versus the product of volume flows for hydrogen and oxygen as motivated in Section~\ref{RS_ModelingTFlame}. The background colors mark which of the operation conditions are assigned to regimes 1 and 2, respectively.}
    \label{RS_TempOverVolFlowWithoutFit}
\end{figure}

In Figure~\ref{RS_TempOverVolFlowWithoutFit}, the calibrated temperatures for single operation points are plotted against the product of volume flows of hydrogen and oxygen. This relationship is investigated under the assumption that the energy released from combustion per time is proportional to the oxygen flow, as all oxygen is consumed in the flame. Furthermore, it is assumed that the flame size is in first order proportional to the total gas volume flow, which is dominated by the hydrogen flow. It is expected that the particles mostly do not pass the most upper flame region due to the gas flow ejected from the burner. Sohn reports that particles fed through the side slot do not melt, whereas particles fed through the burner do \cite{Sohn2023}, which justifies this assumption. Therefore, it is assumed that 
the average temperature acting on particles in the flame region increases with increasing flame size. 
As the objective of this work is the development of a framework that can be extended to more general problems, these simplified assumptions have been used as a first approach to tackle the modeling task. 

In Figure~\ref{RS_TempOverVolFlowWithoutFit}, two overlapping regimes can be identified, one for low products of hydrogen and oxygen flows and one for higher products, accordingly marked by the background color. 
For both regimes, an approximately linear increase of temperature with different slopes is observed. 
For this reason, a first-order temperature model up to a linear term for each regime is established as follows: 

\begin{align}
    & T_{1} (\Dot{V}_{\textrm{H}_2} \cdot \Dot{V}_{\textrm{O}_2}) = k_1+m_1\cdot\Dot{V}_{\textrm{H}_2} \cdot \Dot{V}_{\textrm{O}_2} \ \textrm{\orange{(Regime 1)}},\\
    & T_{2} (\Dot{V}_{\textrm{H}_2} \cdot \Dot{V}_{\textrm{O}_2}) = k_2+m_2\cdot\Dot{V}_{\textrm{H}_2} \cdot \Dot{V}_{\textrm{O}_2}\ {\textrm{\color{green}{(Regime 2)}}},
\end{align}
where $T_1$ and $T_2$ are the PSR temperatures for operating conditions in regimes 1 and 2, respectively, $k_1$ and $m_1$ are the coefficients of the linear function for regime 1 and $k_2$ and $m_2$ are the coefficients of the linear function for regime 2. An overview on which operation points belong to which regime is also given in Table~\ref{RS_RegimeDefinitionTable}. Note, that there are two operation points, which are present in both regimes, making the regimes overlapping. This enables modeling the temperature for operating conditions from the full design space. However, the consistency of the results obtained for operation conditions, whose product of hydrogen and oxygen flow rate lies between the values for these two operation points, has to be examined after calibration. 

\begin{table}[h]
    \centering
        
    \begin{tabular}{ll}
    Regime& Operation points \\
     \hline
       1  & A, B, C, D, E, F, G, H, I, J  \\
        2 & J, I, K, L, M, N, O, P, Q  \\
         \hline
    \end{tabular}
    \caption{Overview on the operation points contained in regimes 1 and 2 identified in the present study.}
    \label{RS_RegimeDefinitionTable}
\end{table}

\subsection{Calibration and uncertainty quantification for the temperature model \label{RS_CaliTModelParams}}

The calibration procedure described in Section~\ref{TB_CaliP} has been used with the default options given in Table~\ref{Method_DefaultCaliOptions} for calibration of $k_1$ and $m_1$ for regime 1 and $k_2$ and $m_2$ for regime 2. Further details specific to this calibration are given in Table~\ref{RS_TModelCali_Options}. 

\begin{table}[h]
    \centering
    \begin{tabular}{ll}
    Parameter & Value\\
    \hline
PCE order (model parameters)  &  1\\
& \\
  Surrogate training regions (CRN response)   & \\
    $k_1$   & $1000 - \SI{1450}{\kelvin}$\\
    $m_1$   & $0.5 - \SI{8}{\kelvin\square\minute\per\square\liter}$\\
    $k_2$   & $1300 - \SI{1550}{\kelvin}$\\
    $m_2$   & $0 - \SI{1}{\kelvin\square\minute\per\square\liter}$\\
         \hline
    \end{tabular}
    \caption{Calibration options used for the calibration of the temperature model described in Section~\ref{RS_ModelingTFlame}, in addition to the default options declared in Table~\ref{Method_DefaultCaliOptions}. }
    \label{RS_TModelCali_Options}
\end{table}

\begin{figure}[h]
    \centering
    \includegraphics[width=0.7\columnwidth]{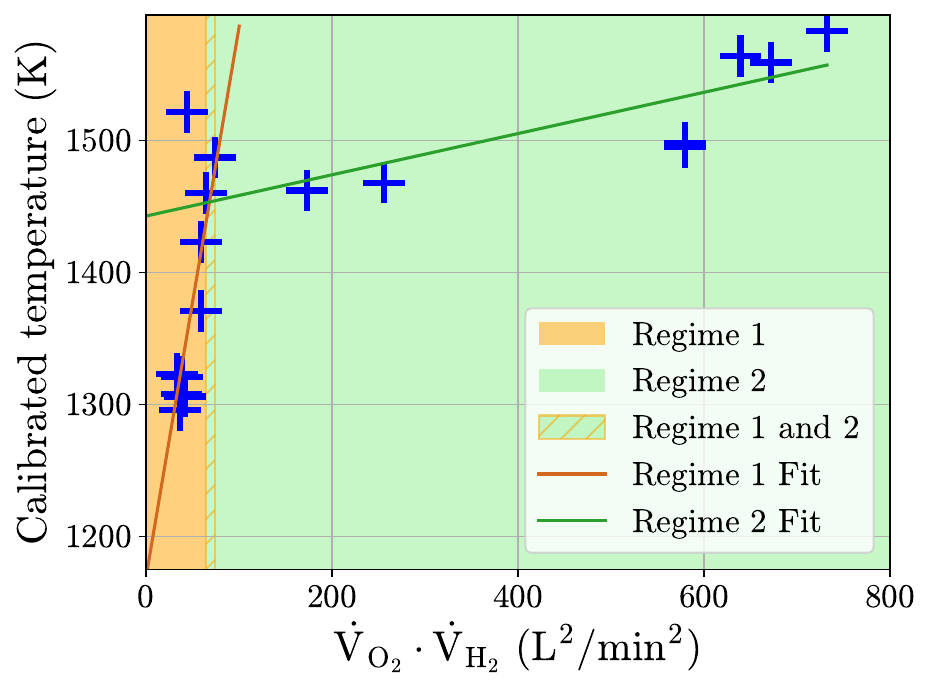}
    \caption{Temperatures calibrated for single operating conditions in Section~\ref{RS_ClassicTCali}, plotted versus the product of volume flows for hydrogen and oxygen as motivated in Section~\ref{RS_ModelingTFlame}. The background colors mark which of the operation conditions are assigned to regimes 1 and 2, respectively. The linear temperature model for each regime, when mean values for the coefficients $k_ 1$, $m_ 1$, $k_2$ and $m_2$ obtained in the calibration are applied, is also shown. Note, that calibration has been conducted with respect to the experimental reduction degrees reported in \cite{Sohn2023} and not directly with respect to the calibrated temperatures for single operating conditions. }
    \label{RS_TempOverVolFlowWithFit}
\end{figure}

In Figure~\ref{RS_TempOverVolFlowWithFit}, the calibrated linear models for the two regimes are shown together with the temperatures calibrated in Section~\ref{RS_ClassicTCali}. It is important to note that calibration has been done with respect to the experimental reduction degrees, not the calibrated temperatures, although these have been used to motivate the model and determine the priors. This is relevant as the sensitivity of the reduction degree to the PSR temperature is different for different operating conditions. 

The calibrated mean values and standard deviations for $k_ 1$, $m_ 1$, $k_2$ and $m_2$ are given in Table~\ref{RS_Resultsk1m1k2m2}. 

\begin{table}[h]
    \centering
    \begin{tabular}{lll}
    \hline
        & Mean& Standard deviation\\
        \hline
     $k_ 1$    & $\SI{1170.16}{\kelvin}$ &$\SI{106.92}{\kelvin}$\\
     $m_ 1$    &$\SI{4.16}{\kelvin\square\minute\per\square\liter}$ &$\SI{1.49}{\kelvin\square\minute\per\square\liter}$\\
     $k_ 2$    & $\SI{1442.73}{\kelvin}$ &$\SI{29.68}{\kelvin}$\\
     $m_2$    & $\SI{0.16}{\kelvin\square\minute\per\square\liter}$ &$\SI{0.044}{\kelvin\square\minute\per\square\liter}$\\
     \hline
    \end{tabular}
\caption{Mean and standard deviation obtained during the calibration of $k_ 1$, $m_ 1$, $k_2$ and $m_2$. }
    \label{RS_Resultsk1m1k2m2}
\end{table}

\begin{figure}[h]
    \centering
    \begin{subfigure}[c]{0.49\textwidth}
        \includegraphics[width=0.99\textwidth]{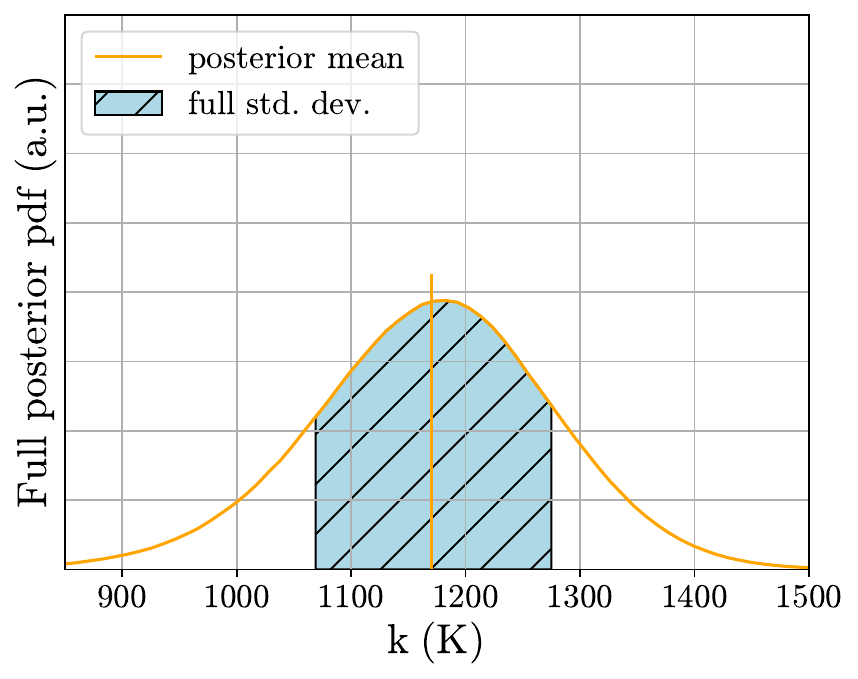}
        \caption{parameter $k_1$}
    \end{subfigure}
        \begin{subfigure}[c]{0.49\textwidth}
        \includegraphics[width=0.99\textwidth]{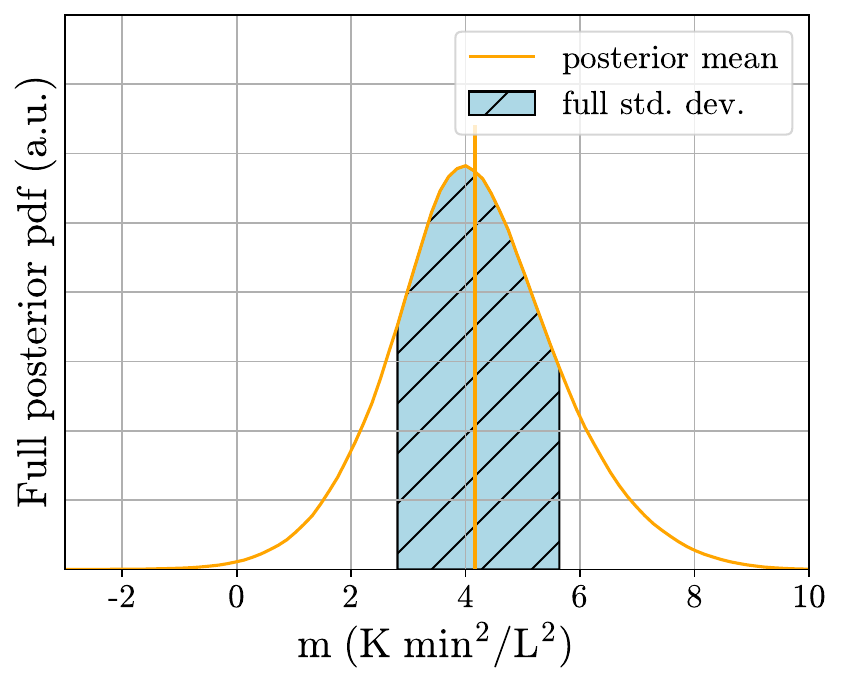}
        \caption{parameter $m_1$}
    \end{subfigure}
    \caption{Full posterior PDF for the model parameters $k_1$ and $m_1$ for the temperature model in regime 1. }
    \label{RS_k1m1PDF}
\end{figure}

In Figure~\ref{RS_k1m1PDF}, the full posterior PDFs for the model parameters of regime 1, $k_1$ and $m_1$ are shown. It is observed that both PDFs are rather broad, with $m_1$ even covering negative values, although with very low PDF values. This can be understood from Figure~\ref{RS_TempOverVolFlowWithoutFit}, as the temperature increase is only approximately linear and not even monotone, with especially one outlier, OP F, for which the calibrated temperature from Section~\ref{RS_ClassicTCali} has been a lot higher than proposed by the linear model. This reveals, that there is potential to further enhance the applied model. However, for these considerations also the results for the reduction degrees have to be taken into account. 

\begin{figure}[h]
    \centering
\begin{subfigure}[c]{0.49\textwidth}
    \centering
    \includegraphics[width=0.99\textwidth]{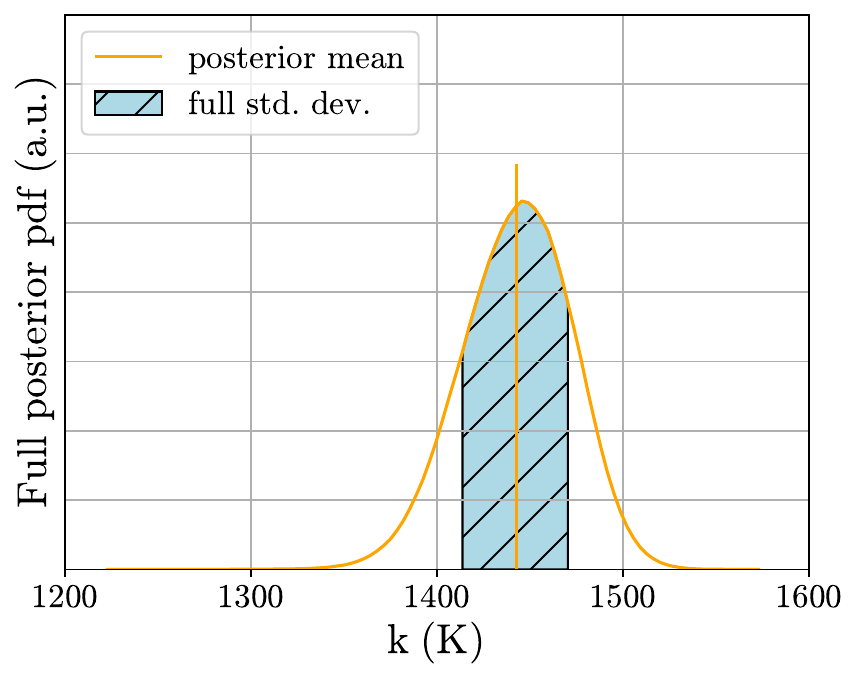}
    \caption{parameter $k_2$}
    \end{subfigure}
            \begin{subfigure}[c]{0.49\textwidth}
            \centering
        \includegraphics[width=0.99\textwidth]{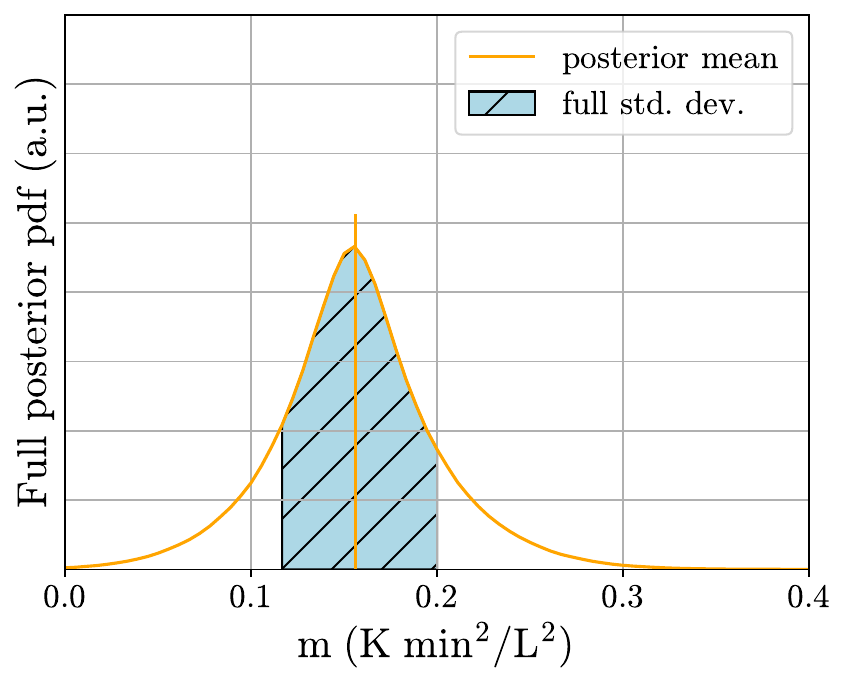}
        \caption{parameter $m_2$}
    \end{subfigure}
    \caption{Full posterior PDF for the model parameters $k_2$ and $m_2$ for the temperature model in regime 2. }
    \label{RS_k2m2PDF}
\end{figure}

In Figure~\ref{RS_k2m2PDF}, the full posterior PDFs for parameters $k_2$ and $m_2$ are presented. 
For $k_2$, the width of the curve, i. e. the standard deviation is much lower than for $k_1$, whereas the curve width for $m_2$ is similar as for $m_1$. The reduced spread for $k_2$ compared to $k_1$ can be due to the slope of temperature increase being a lot lower in regime 2 compared to regime 1.

An overview of the full posterior predictions for the reduction degree for all operation points in regime 1 is given in Appendix~\ref{AP_CaliResultsAll}, Figures~\ref{AP_Reg1_FullPostPredPDFsA-F} and~\ref{AP_Reg1_FullPostPredPDFsG-R}. It is observed that the PDF curves differ in width and also on how well the mean prediction fits the experimental value. The same holds for regime 2, for which the full posterior predictions are shown in Figure~\ref{AP_Reg2_ALLOPS_FullPostPredPDFsI-N} and~\ref{AP_Reg2_ALLOPS_FullPostPredPDFsO-R}. 
As the ABC method has been used for likelihood evaluation (see Section~\ref{TB_Likelihood}), it is expected that the deviation between the mean prediction and the experimental value is on average in the order of the standard deviation, at least for samples of the parameter vector $\bm{\tilde{\alpha}}$ with a high posterior value. 
In Table~\ref{AP_Reg1_Pred_MeanExpDev2Stddev}, the experimental reduction degrees reported by Sohn \cite{Sohn2023}, the mean calibrated values, deviation between the mean prediction and experimental value as well as twice the standard deviation, as width of the $1\sigma$-interval around the mean, are listed for all operation points used in regime 1. For the deviation and (twice) the standard deviation, also average values for all points
are given, respectively. It is observed that the average (single) standard deviation is approximately the average deviation for the seen points from regime 1. 
Also for regime 2, the average standard deviation of the seen points fits quite well to the average deviation between mean predictions and experimental values, with the average standard deviation being even a bit higher than the averaged deviation \footnote{Note, that the ABC condition is only influencing the likelihood for each single sample $\bm{\tilde\alpha}$. However, both the mean standard deviation for single samples as well as the posterior standard deviation of the mean predictions contribute to the total standard deviation, which means the condition should be approximately fulfilled for the full posterior distributions as well. }. 

A further analysis can be conducted with the help of Tables~\ref{AP_Reg1_Pred_MeanExpDev2Stddev} and~\ref{AP_Reg2_ALLOPS_Pred_MeanExpDev2Stddev} given in Appendix~\ref{AP_CaliResultsAll}. For regime 1, the average deviation between the mean prediction of the reduction degree and the experimental value is 0.0142, with the maximum deviation being 0.049 for operation point F, which has already been identified as an outlier. Regarding that a simple CRN model has been used together with a linear temperature model, the predictive capacity of the model is quite good. For regime 2, the average deviation being 0.0242 is higher than for regime 1, but still very low. The maximum deviation in regime 2 is 0.042 for OPs M and Q. The average deviation over all operation points used in the present study, OP I and J being considered for both regimes, is 0.020, which is the least experimental uncertainty expected for the data, see Section~\ref{RS_DataError}. For OP F, the experimentally obtained reduction degree is the very upper bound of the prediction interval, for all other operation points, the experimental value has a non-zero PDF value. 
This way, it can be stated that a CRN model of relatively low complexity has been found, which is able to produce acceptable predictions of the reduction degrees obtained in the laboratory reduction reactor developed by Sohn and co-workers \cite{Sohn2023}. 

However, also uncertainty quantification is an important part of the prediction, which here means quantification of the model-induced uncertainty, which has been embedded in the model parameters and determined in the calibration along the mean values. The standard deviation of the predictions is an important measure for the quantification of the model error, especially since the ABC method has been used for the likelihood evaluation of samples. %

Finally, it is to be analysed how meaningful the obtained results are, i.e. how large the uncertainty is relative to the absolute values of the mean prediction and the size of the image set of the CRN model, which here is the interval $\left[0,1\right]$. 

As the ABC method has been used for likelihood evaluation, the standard deviation of the posterior curves is on average a measure for the deviation between the mean prediction and the "true" data, i.e. the model error. With an average value of 0.017 for regime 1, and 0.027 for regime 2, the relative model error is small compared to the absolute predictions for the reduction degree. However, it is observed that the standard deviation mostly increases with decreasing mean prediction, meaning that the relative model error is higher for operating conditions leading to lower reduction degrees. This is understandable as the rate of change for the reduction degree, cf. Equation~\eqref{TC_ReductionDegreeRateEquation}, for given operating conditions, decreases with increasing reduction degree, such that also the partial derivative with respect to the temperature is higher for lower values of the reduction degree. For regime 1, the highest relative model error (= standard deviation) for OP J is 3.6\%, whereas the lowest for OP F is 1.2\%, for regime 2 the highest relative model error is 10.2\% for OP Q, whereas the lowest value is 1.2\% for OP K. 

Another point to consider is the gain of information obtained from the calibrated CRN model with model error quantification. In the present case, this concerns the region within the interval $\left[0,1\right]$ into which the prediction of the reduction degree can be limited \footnote{A trivial example with no gain of information would be if for all or some of the operation conditions, the posterior prediction PDF would be uniform on $\left[0,1\right]$.}. However, as the posterior prediction PDFs are strongly non-uniform, there is no unique definition of this region available and it also depends on the application. For instance, the design purpose might be output maximization, or highest robustness. 

If using the result, where possible, i.e. computationally tractable, the full PDF information on the prediction should be employed. Other than this, the integrated probability value of a certain interval can be examined, i.e. the one-, two- and three-standard-deviations-intervals ($1\sigma$-, $2\sigma$-, $3\sigma$-intervals) around the mean. For regime 1, this analysis has revealed that on average, the truth value is within the $1\sigma$-interval with a probability of 0.665, in the $2\sigma$-interval with a probability of 0.947 and in a $3\sigma$-interval with a probability of 0.992, with a low scatter among the operation points. For regime 2, it has been found that the truth value is within the $1\sigma$-interval with a probability of 0.646, in the $2\sigma$-interval with a probability of 0.962 and in a $3\sigma$-interval with a probability of 0.999, also with a low scatter among the operation points. Interestingly, these values are quite similar to the well-known values for Gaussian curves. 

As has been seen for the single standard deviations, there is a strong increase in the width of the $1\sigma$, $2\sigma$ and $3\sigma$-intervals with decreasing mean prediction for the reduction degree. Independently of the exact quantity of interest 
it can be concluded that there is a clear gain of information through the calibrated CRN model, especially for those operating conditions, for which high reduction degrees are obtained. According to Sohn, these points are the most relevant for the design process, anyway \cite{Sohn2023}. Quantities of interest could for instance be the mean prediction and its standard deviation, or the lower bound of the prediction. 
Still, in order to obtain a useful model on the full design space, a refinement of the model such that the model error becomes lower for those operating conditions leading to intermediate reduction degrees, while keeping it low for the high ones, would be beneficial. 

Before discussing the obtained results in the light of the overall objective of the present work, there will be further comments on the overlapping of the regimes and the role of the data error, as well as possible model refinements. 

\subsection{Comments on the overlapping of regime 1 and 2}

As has been introduced in Section~\ref{RS_ModelingTFlame}, the two regimes for the temperature modeling have been defined such that there is an overlap between them within the design space. The operation points I and J (see Table~\ref{TC_OPTable}) both belong to regime 1 and 2 and the same shall hold for any operation point whose product of hydrogen and oxygen is between those of OP I and J. 

\begin{figure}[h!]
    \centering
    \begin{subfigure}[t]{0.33\textwidth}
    \centering
        \includegraphics[width=0.99\columnwidth]{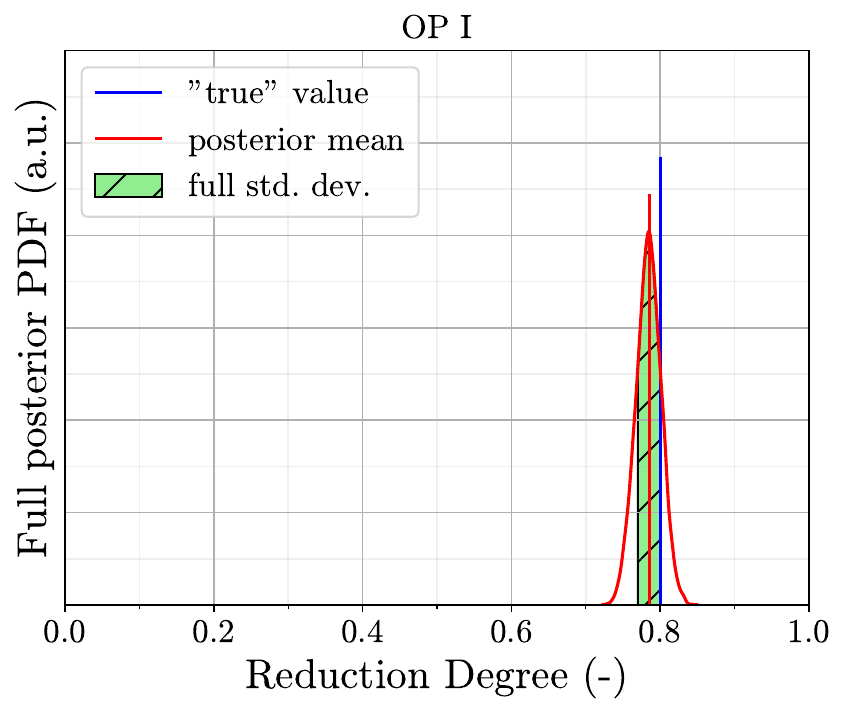}
        \caption{Regime 1, OP I.}
    \end{subfigure}
        \begin{subfigure}[t]{0.33\textwidth}
            \centering
        \includegraphics[width=0.99\columnwidth]{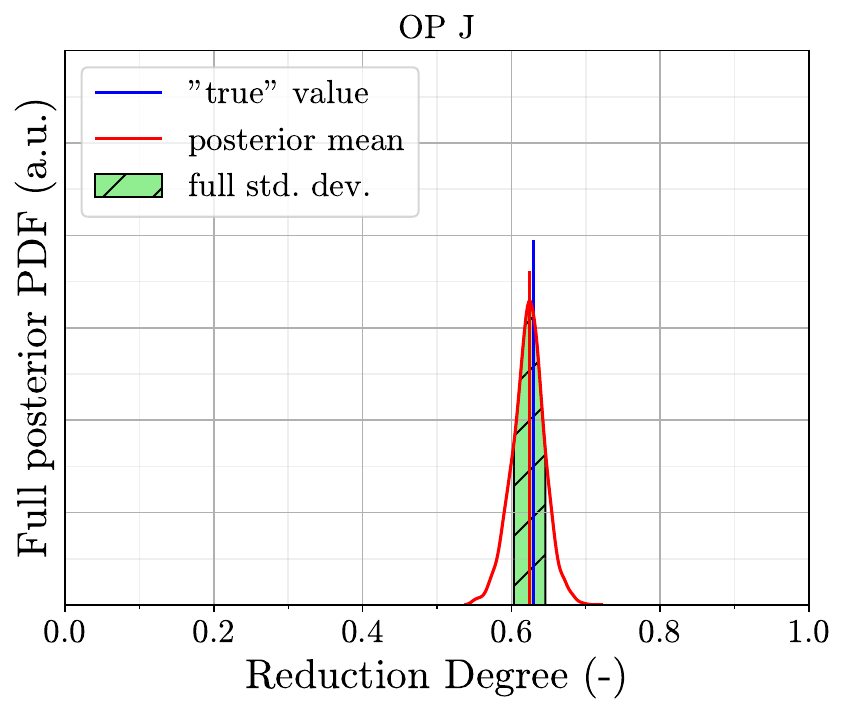}
        \caption{Regime 1, OP J.}
    \end{subfigure}
        \begin{subfigure}[t]{0.33\textwidth}
            \centering
        \includegraphics[width=0.99\columnwidth]{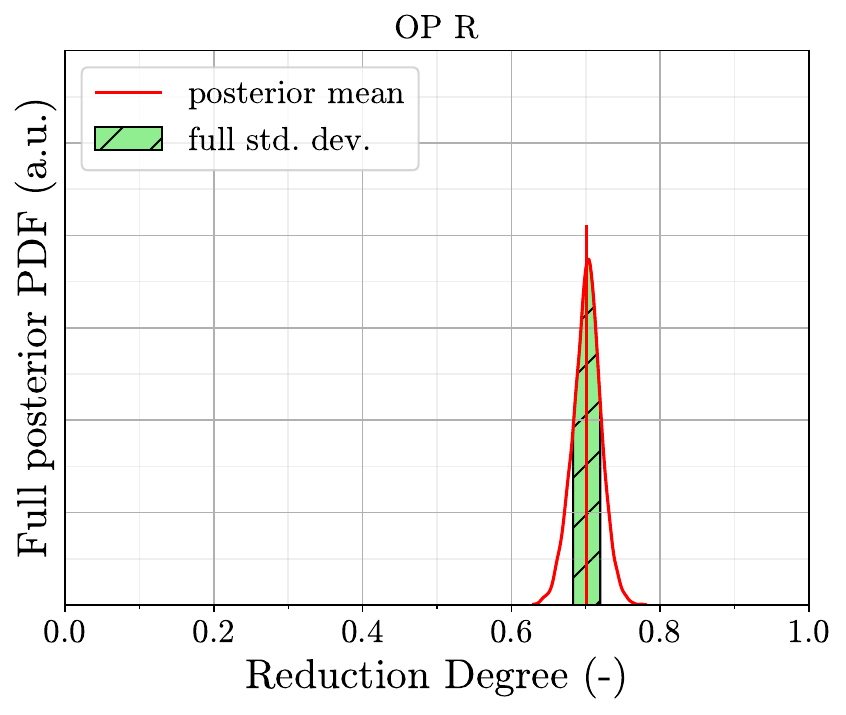}
        \caption{Regime 1, OP R.}
    \end{subfigure}
        \begin{subfigure}[t]{0.33\textwidth}
            \centering
        \includegraphics[width=0.99\columnwidth]{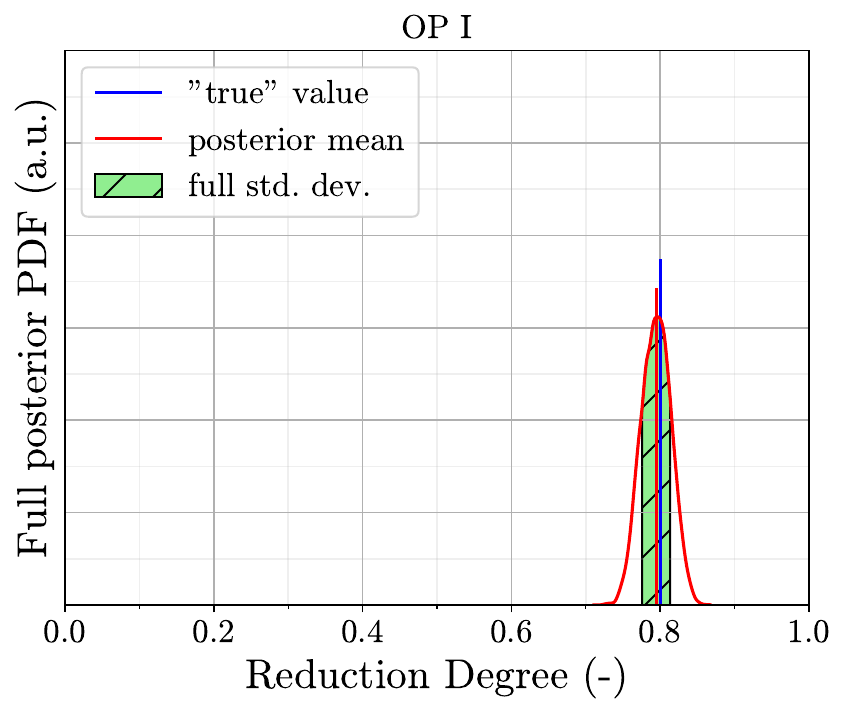}
        \caption{Regime 2, OP I. }
    \end{subfigure}
        \begin{subfigure}[t]{0.33\textwidth}
            \centering
        \includegraphics[width=0.99\columnwidth]{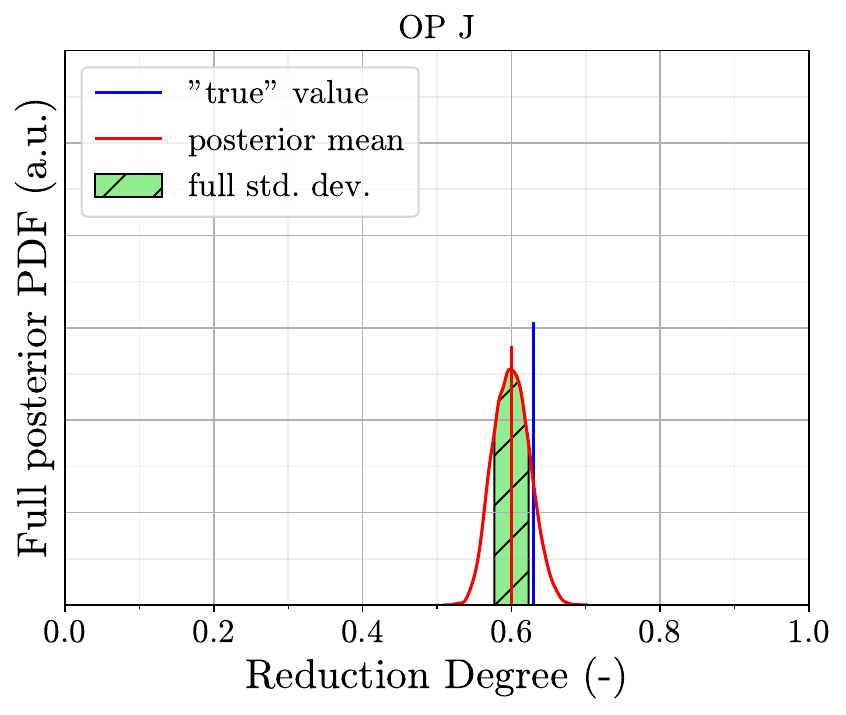}
        \caption{Regime 2, OP J. }
    \end{subfigure}
        \begin{subfigure}[t]{0.33\textwidth}
            \centering
        \includegraphics[width=0.99\columnwidth]{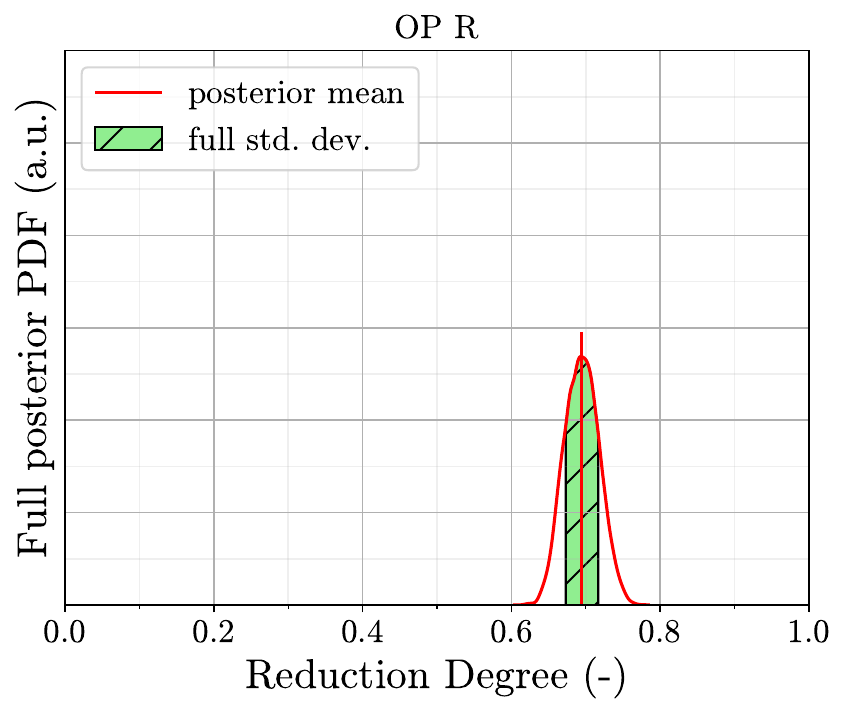}
        \subcaption{Regime 2, OP R. }
    \end{subfigure}
    \caption{Comparison of the results for regime 1 and 2 for the operation points being present in both regimes. }
    \label{RS_OverlappingResults}
\end{figure}

For this reason, for these points there are two concurring solutions from regime 1 and 2, whose consistency therefore should be examined. This is why the CRN predictions for a further operation point out of the overlapping region (OP R) have been evaluated. The results for OPs I, J and R for regime 1 and 2 are shown in Figure~\ref{RS_OverlappingResults}. 
All mean predictions for one regime are still within one standard deviation of the mean result for the other regime, except fro OP J, for which the prediction in regime 2 is outside the $1\sigma$-region of regime 1. For OP R, the mean predictions for regime 1 and 2 are quite close to each other. For OP J, the prediction of regime 1 is better than the one of regime 2.   For OP R, no such statement on the mean prediction is possible as there is no validation data for this point available. 

All together, it can be stated that the results in the overlapping region are not entirely consistent but the deviations are still acceptable. Still, a redefinition of the boundary between the two regimes can be considered. E.g., it may be possible to define just one operation point as the boundary, where the mean predictions are very close. In doing so, not only the mean prediction but also the consistency between the standard deviations, which quantify the model error, is to be considered.

\subsection{Role of the data error \label{RS_DataError}} 

As introduced in Section~\ref{TB_Model}, next to the model error, there can also be a data error, which is the deviation between the unknown "truth" and the actual data. Especially in the scope of experiments the data error is an omnipresent concept. There are several ways to account for the data error. There can be information from the experiment, e.g. on the reproducibility of data, but the data error can also be inferred within the calibration procedure, which can lead to a modified result \cite{Sargsyan2019}. 

For the given case, there is only limited information on the data error available to the authors. A data reproducibility of $\pm 2\%$ is reported by Sohn for the comparison between different methods of the evaluation of the reduction degree \cite{Sohn2023}. Thus, this can be seen as a lower bound for the data error in the given case. 

For some of the operation points used (see Table~\ref{TC_OPTable}), there are also results of CFD simulations available, see \cite{Sohn2023}. Especially for lower reduction degrees there partially is a strong deviation between CFD and experimental results with the maximal absolute difference for the used operation points being 0.12 for OP J. This leads to the presumption that the overall data error might be higher than $\pm 2\%$. 

To summarize, the magnitude of the data error is not known very well for the given case, and further examinations need to be done on that issue in future work. 

\subsection{Possible model improvements \label{RS_OP_F}}

The analysis conducted in the above sections has shown that the proposed model on average has a high quality in correctly  predicting the experimental values for the reduction degree obtained in the laboratory reactor under investigation. However, two aspects for possible future enhancement shall be discussed separately. 

Firstly, the highest deviation between experimental and predicted value is 0.049 for OP F. This is the only operation point tested, where the experimental value is only at the very upper bound of the prediction. As has been presented in Section~\ref{RS_DataError}, there is a non-negligible data uncertainty, which can partially explain that outliers exist. Still, there are indicators for possible improvements which especially hold for OP F, and furthermore, Sohn states that those operation conditions leading to very high reduction degrees are the most important for the design process of reactors for flash ironmaking \cite{Sohn2023}. In Section~\ref{SetUpCRN}, the ideal gas approach used to calculate the residence time for the present study has been explained. For comparison with the values provided by Sohn, a constant temperature of $\SI{1448}{\kelvin}$ throughout the full CRN has been assumed. 
For OP F, our calculation leads to a $\SI{1.4}{s}$ shorter particle residence time compared to the reference. This is the highest deviation among the operation points used in this work. 
Increasing the residence time would lead to lower temperature values in the calibration. Moreover, the particle density decreases during reduction, which Sohn and co-workers have considered in their CFD computations \cite{Fan2016}. This has not been integrated into the present model yet and would increase the residence time especially for high reduction degrees. It is not known to the authors if this has already been considered during Sohn's calculations (for experiments) cited above. 

Secondly, there might be an offset in the calibrated PSR temperatures compared to the actual particle temperature in the PSR region. Fan has published a particle temperature profile for the operation point that is denoted as OP N in the present work \cite{Fan2019}, revealing lower particle temperatures compared to our calibration for that OP. However, it has to be considered that according to \cite{Fan2019}, a different approach for the equilibrium constant of the magnetite reduction reaction has been used, leading to differing values from those determined experimentally. This can have an influence on the calculated temperatures, and on the other hand, using those equilibrium constants from \cite{Fan2019} would probably have a significant influence on the calibrated temperatures. However, this does not necessarily decrease the quality of the prediction for the reduction degree as long as the equilibrium constant (and all other assumptions) are consistent between model calibration and the usage of the model. Still, the physical interpretation of the absolute temperature values might be affected by this and the influence of the equilibrium constant on the values of the parameters in the temperature model should be investigated in future work.

\section{Discussion}\label{Discussion}

As has been introduced in Section~\ref{Introduction}, the motivation of the present work is the question to which extent CRNs can be used as a scale bridge within a model hierarchy spanning a vast range of scales, in the context of developing an innovative metal-based energy circular economy. 
Therefore, CRN models have to be valid on the full design space of operating conditions and not just for those operating conditions, where also data from experiments or computational fluid dynamics simulations exist. Also, the transferability of CRN models among scales is an important point to consider. As CRNs are models of reduced complexity and therefore, reduced accuracy, the quantification of uncertainties rising from the model itself can be helpful to find a universal model and to know where to improve it. 

In the present work, a first step has been achieved in finding and calibrating a CRN model to describe a laboratory reactor developed by Sohn and co-workers \cite{Sohn2023} on the full design space. 
This has been achieved by combining model-to-model-calibration methods \cite{Sargsyan2015,Sargsyan2019} with CRNs, as suggested by Saverese \textit{et al.} \cite{Savarese2023b}. 
In the present case, the introduction of \textit{meta} parameters to the CRN model has been shown to be a useful concept for defining a CRN model on a range of operating conditions while keeping a simple CRN structure. Also the partitioning of the design space into two regimes has been helpful, since this way it has been possible to use a linear, that is, "first order" model for the parametrization of the PSR temperature. This is especially helpful if the underlying phenomena are not known to a sufficient extent or not enough data is available for a deeper analysis, which is often the case for high-fidelity models that are difficult to evaluate. The simple model has been found to perform reasonably well, in terms of the low deviation of the prediction from the experimental data, and regarding that the relative model error is low for most operation points. This holds especially for those operation points, for which elevated reduction degrees are obtained, which are the most relevant ones for the design purpose, see also \cite{Sohn2023}. 

One open issue regarding the present case is a lack of data for intermediate values of $\Dot{V}_{H_2}\cdot\Dot{V}_{O_2}$ in regime 2, see Figure~\ref{RS_TempOverVolFlowWithoutFit}. Therefore, it will be beneficial to enhance the underlying temperature model, which can for example be achieved by explicitly modeling the flame. This could also lead to smaller model errors in the calibrated parameters themselves, especially in regime 2. When comparing the standard deviations of the calibrated parameters to those of the model prediction, it seems that the PSR temperature might not be the most sensitive parameter for the reduction degree. Indeed, it is plausible that the strongest driving force of the reduction is the amount of excess hydrogen, which is also elaborately discussed by Sohn \cite{Sohn2023}. 
This can be further examined by employing sensitivity analysis. 
Note, that the excess hydrogen is mostly controlled by the operation conditions, i.e. the hydrogen and oxygen flow, which are model inputs to the CRN model, and thus is not a model parameter of the CRN itself. 

These points reveal that a generalization of the framework to other, especially more complex cases, will be a challenge for future work. For example, also within the scope of the present research on metal-fueled energy cycles, there are application cases with more complex reactor geometries, chemical reaction mechanisms, and potentially less data available, as the higher the complexity, the more difficult the evaluation of high-fidelity models. Moreover, for these more complex situations, probably there will be a need to apply more advanced methods of uncertainty qunatification, i.e. for efficient sampling. 
Still, the present work is a first step that will be further extended. 

Another important point to make is on scaling. In the present work, the calibration framework has been tested for a laboratory-scale reactor. For being useful in the development of a large-scale application, the modeling approach needs to be transferred to industrial scales. In the case of flash ironmaking, experimental data for several pilot scales and simulation data from pilot to industrial scale reactors are available \cite{Sohn2023}. Therefore, the model developed for the laboratory reactor will be extended to these scales in a next step. However, in the scope of design and optimization problems, it should be considered to incorporate geometry and size as variables themselves, making them accessible to optimization and efficiency analysis along with the other relevant parameters. 

\section{Conclusion and Outlook\label{Conclusion + Outlook}}

Within the present work, a laboratory reactor for innovative flash ironmaking, for which data is available in the literature \cite{Sohn2023}, has been modeled by a chemical reactor network. This model has been developed based on secondary data from experiments and computational fluid dynamics simulations from \cite{Sohn2023} with the objective of creating a valid model on the full design space, while data is only available on a discrete, finite subset. Uncertain parameters have been calibrated along with quantification of uncertainties resulting from the inaccuracy of the model, following an approach suggested by Savarese \textit{et al.} \cite{Savarese2023b} based on the work by Sargsyan \textit{et al.} \cite{Sargsyan2015,Sargsyan2019}. A simple CRN structure consisting of a turbulent and a straight flow pattern, where particle slip is taken into account, has been identified to work very well in combination with a \textit{meta} model relating the temperature in the flame region with the operating conditions. 

The long-term objective of this work is using CRN models within large-scale optimization and efficiency analyses, while incorporating information from the nano- to the mesoscale, in the development of large-scale applications, especially in the scope of the current development of a metal-fueled energy circular economy \cite{bergthorson2015,janicka2023}. 
The next steps include enhancements of the model described in this article, among them it will be investigated how to improve the flame modeling, and possibly alternative approaches for the temperature model. Also, including more CRN model parameters being uncertain, potentially accompanied by sensitivity analysis, will be considered if it shows to be beneficial. There will be further investigation of model error convergence and its stability among the design space and of the role of data errors and their quantification. As discussed in Section~\ref{Discussion}, upscaling will subsequently be an important step for reaching the long-term objective.

\section*{CRediT authorship contribution statement}

\textbf{L. Gossel:} Conceptualization, Methodology, Software, Validation, Formal analysis, Investigation, Data curation, Writing - original draft, Writing - review \& editing, Visualization, 
\textbf{E. Corbean:} Conceptualization, Methodology, Writing - original draft, Writing - review \& editing,
\textbf{S. Dübal:} Conceptualization, Writing - original draft, Writing - review \& editing, Visualization, 
\textbf{P. Brand:} Software, Investigation, 
\textbf{M. Fricke:} Writing - review \& editing, Supervision, Project management, 
\textbf{H. Nicolai:} Writing - review \& editing, Supervision, Project management, 
\textbf{C. Hasse:} Conceptualization, Supervision, Project management, Funding acquisition, 
\textbf{S. Hartl:} Supervision, Project management, Funding acquisition, 
\textbf{S. Ulbrich:} Supervision, Project management, Funding acquisition, 
\textbf{D. Bothe:} Resources, Writing - review \& editing, Supervision, Project management, Funding acquisition. 

\section*{Acknowledgements}
Funded by the Hessian Ministry of Higher Education, Research, Science and the Arts - cluster project Clean Circles. The authors would like to thank %
Prof. Tiziano Faravelli and 
Prof. Alessandro Stagni for the fruitful discussions and advice regarding CRN modeling and for providing the NetSMOKE framework, 
and Jannik Neumann for the valuable exchange on the integration of technologies for iron oxide reduction. 
\vspace{-0.1 cm}
\section*{Appendix}
\vspace{-0.3 cm}
\appendix
\counterwithin{figure}{section}
\counterwithin{table}{section}
\renewcommand\thefigure{\thesection\arabic{figure}}
\renewcommand\thetable{\thesection\arabic{table}}
\section{Calibration results \label{AP_CaliResultsAll}}
\vspace{-0.3 cm}
\subsection{Calibration of PSR temperature for single operation points}
In Table~\ref{AP_ClassicTCali}, the results of the PSR temperature calibration for single operation points are listed. 
\begin{table}[h]
    \centering
    \begin{tabular}{cc}
    Operation point&Calibrated temperature\\
    \hline
         A & 1323 K \\
         B & 1296 K \\
         C & 1308 K \\
         D & 1321 K \\
         E & 1306 K \\
         F & 1521.5 K \\
         G & 1423 K \\
         H & 1371 K \\
         I & 1460 K \\
         J & 1487 K \\
         K & 1462 K \\
         L & 1468 K \\
         M & 1495 K \\
         N & 1498 K \\
         O & 1564 K \\
         P & 1559 K \\
         Q & 1583 K \\
    \hline
    \end{tabular} 
    \caption{Results of the classical PSR temperature calibration as presented in Section~\ref{RS_ClassicTCali}.}
    \label{AP_ClassicTCali}
\end{table}
\vspace{- 0.8cm}
\subsection{Calibration of model parameters for PSR temperature in regime 1}
\begin{figure}[h]
    \centering
    \begin{subfigure}[b]{0.49\textwidth}
        \centering
        \includegraphics[width = \textwidth]{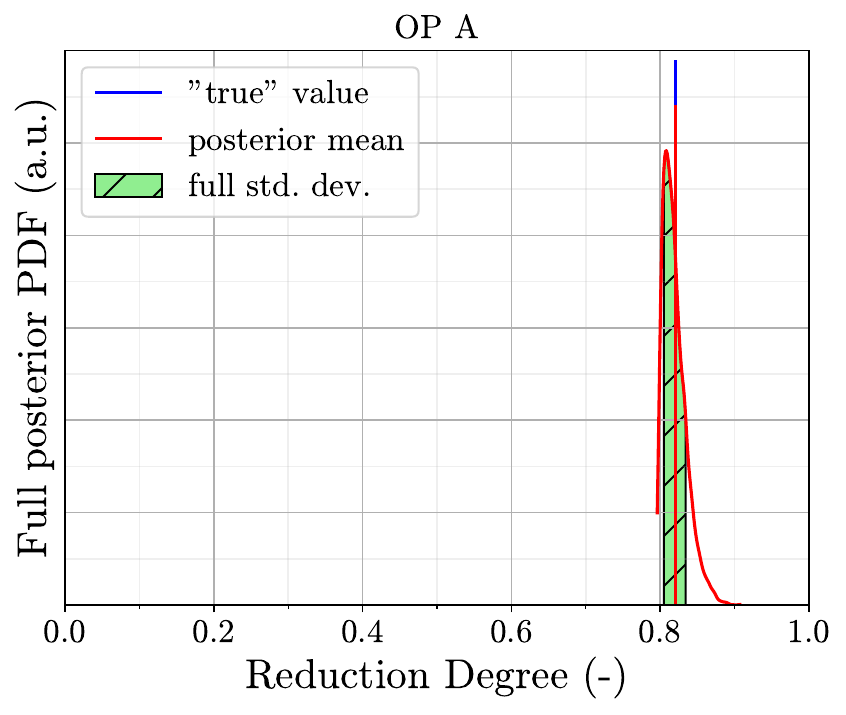}
        \caption{Full posterior prediction for OP A.}
    \end{subfigure}
        \begin{subfigure}[b]{0.49\textwidth}
        \centering
        \includegraphics[width = \textwidth]{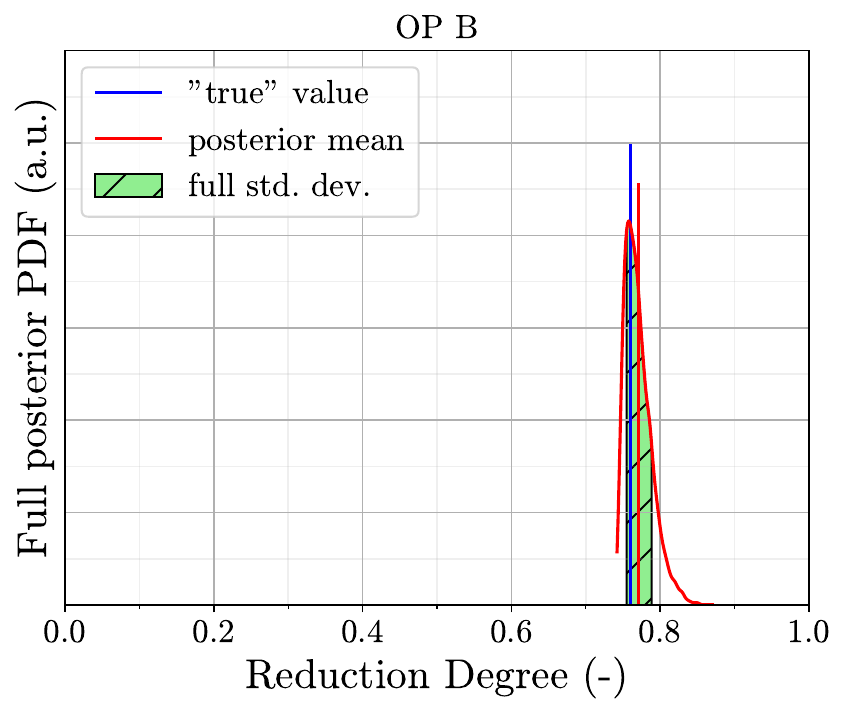}
        \caption{Full posterior prediction for OP B.}
    \end{subfigure}
        \begin{subfigure}[b]{0.49\textwidth}
        \centering
        \includegraphics[width = \textwidth]{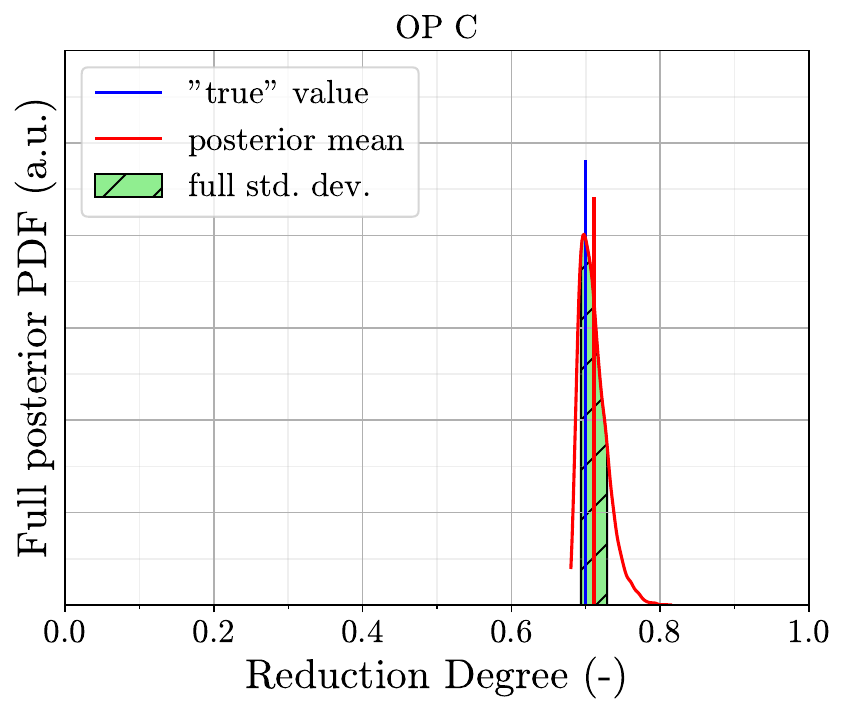}
        \caption{Full posterior prediction for OP C.}
    \end{subfigure}
        \begin{subfigure}[b]{0.49\textwidth}
        \centering
        \includegraphics[width = \textwidth]{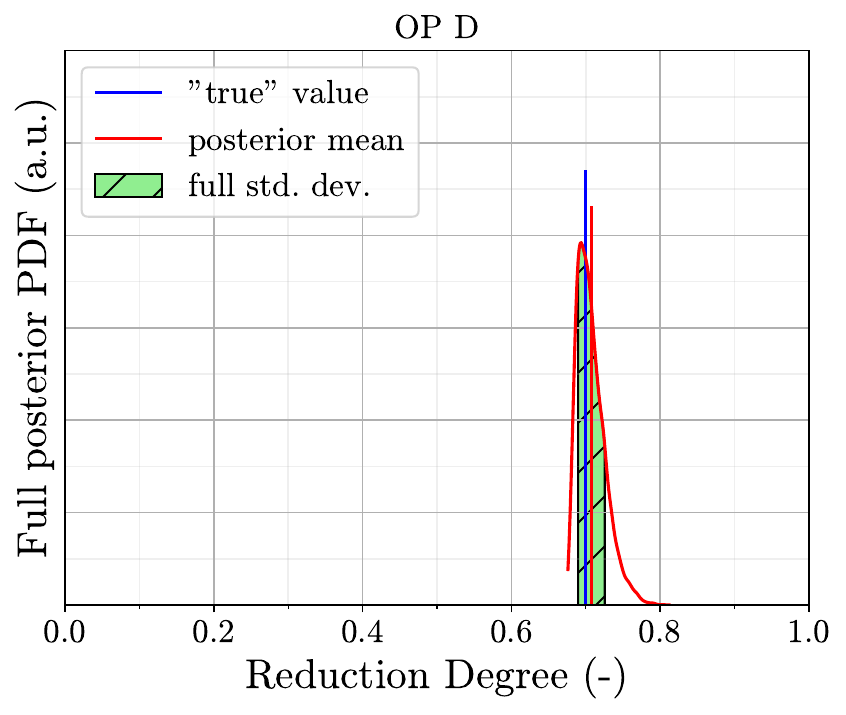}
        \caption{Full posterior prediction for OP D.}
    \end{subfigure}
            \begin{subfigure}[b]{0.49\textwidth}
        \centering
        \includegraphics[width = \textwidth]{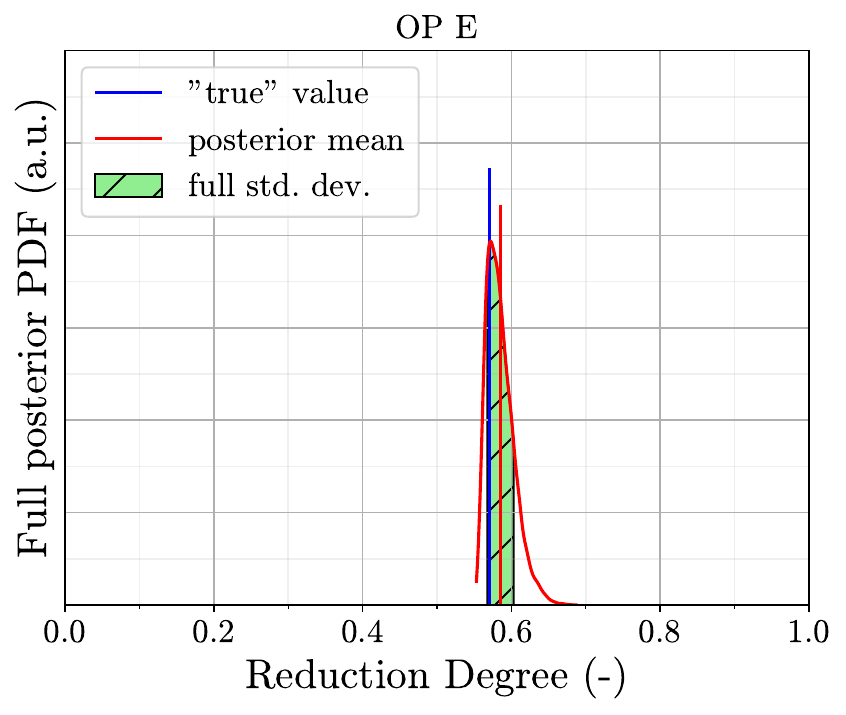}
        \caption{Full posterior prediction for OP E.}
    \end{subfigure}
            \begin{subfigure}[b]{0.49\textwidth}
        \centering
        \includegraphics[width = \textwidth]{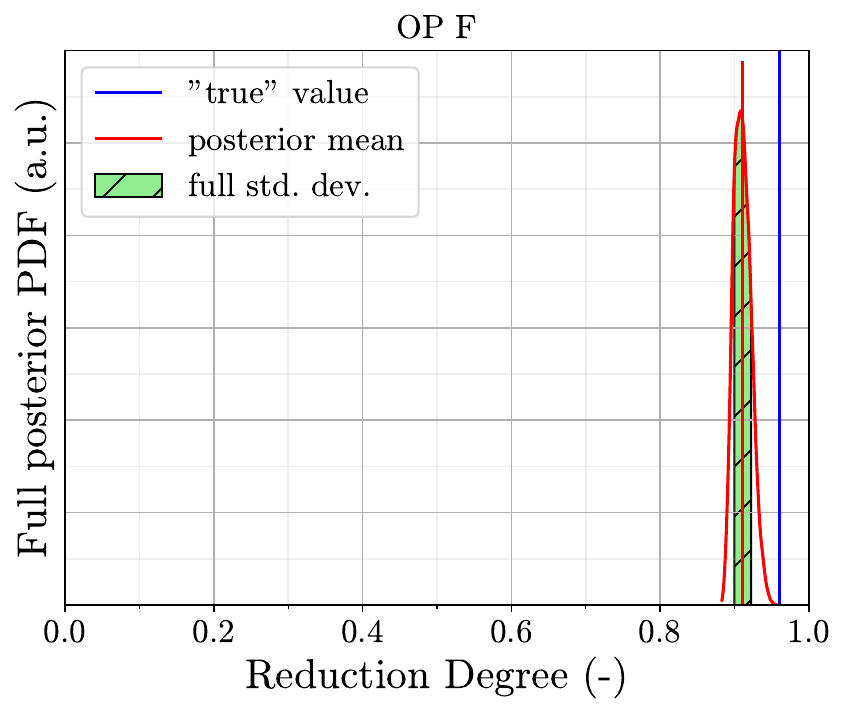}
        \caption{Full posterior prediction for OP F.}
    \end{subfigure}
    \caption{Overview of full posterior predictions for regime 1 and OP A to F. }
    \label{AP_Reg1_FullPostPredPDFsA-F}
    \end{figure}

\begin{figure}[h]
    \centering
                \begin{subfigure}[b]{0.49\textwidth}
        \centering
        \includegraphics[width = \textwidth]{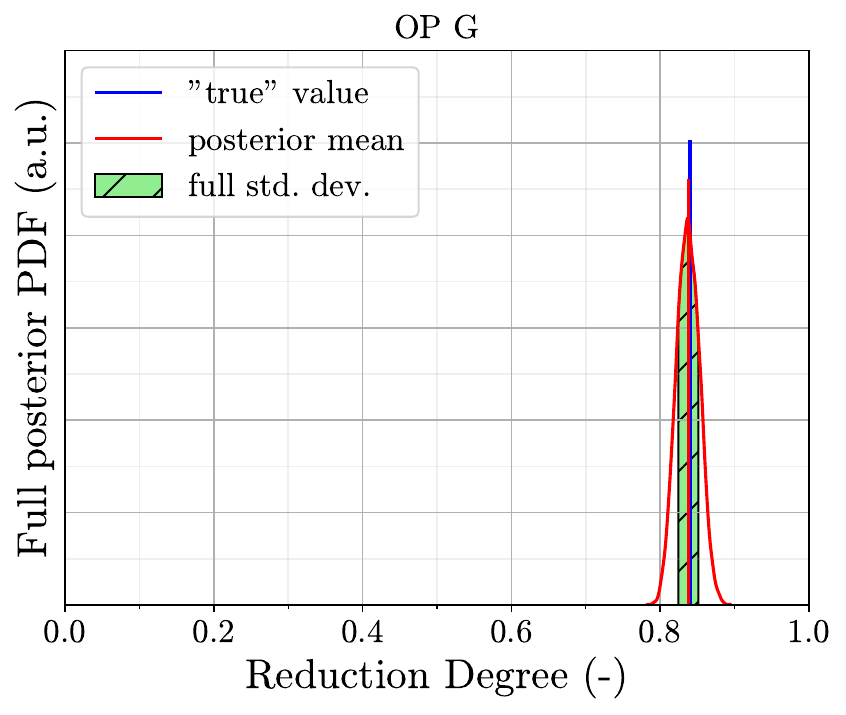}
        \caption{Full posterior prediction for OP G.}
    \end{subfigure}
                \begin{subfigure}[b]{0.49\textwidth}
        \centering
        \includegraphics[width = \textwidth]{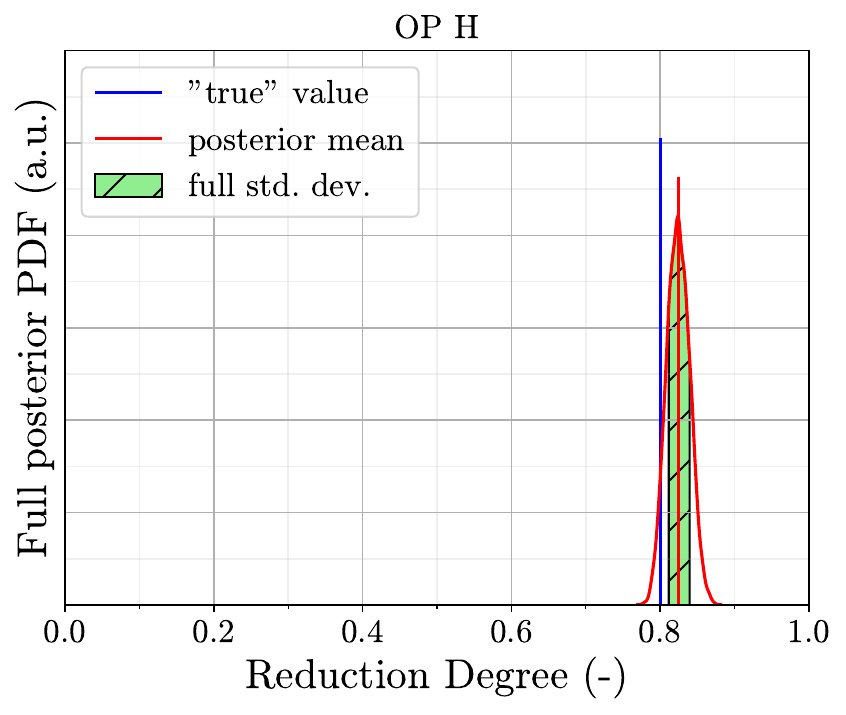}
        \caption{Full posterior prediction for OP H.}
    \end{subfigure}
                    \begin{subfigure}[b]{0.49\textwidth}
        \centering
        \includegraphics[width = \textwidth]{Figures/FiguresRD/FullPostPred_Reg1_OP8.pdf}
        \caption{Full posterior prediction for OP I.}
    \end{subfigure}
                    \begin{subfigure}[b]{0.49\textwidth}
        \centering
        \includegraphics[width = \textwidth]{Figures/FiguresRD/FullPostPred_Reg1_OP9.pdf}
        \caption{Full posterior prediction for OP J.}
    \end{subfigure}
        \begin{subfigure}[t]{0.49\textwidth}
        \centering
        \includegraphics[width = \textwidth]{Figures/FiguresRD/FullPostPred_Reg1_OP17.pdf}
        \caption{Full posterior prediction for OP R (unseen, no validation data available).}
    \end{subfigure}
    \caption{Overview of full posterior predictions for regime 1 and OP G-J and R. }
    \label{AP_Reg1_FullPostPredPDFsG-R}
\end{figure}

In Table~\ref{AP_Reg1_Pred_MeanExpDev2Stddev}, the experimental reduction degrees reported by Sohn \cite{Sohn2023} are compared to the predictions of the CRN model developed within the present work, along with the deviation between the latter two values and width of the $1\sigma$-interval of the prediction, for the operation points of regime 1. 
In Figures~\ref{AP_Reg1_FullPostPredPDFsA-F} and~\ref{AP_Reg1_FullPostPredPDFsG-R}, the full posterior PDFs of the predictions of the reduction degree obtained from the calibrated CRN model for regime 1 are shown. 

\begin{table}[h]
    \centering
    \begin{tabular}{ccccc}
    Operation point&Experimental value &Mean prediction & Deviation &$\sigma\ /\ 2\cdot \sigma$\\ 
    \hline
    \hline
       A   & 0.82&0.820&0.0 &0.015/0.030\\
       B   & 0.76&0.771&0.011 &0.018/0.035\\
        C & 0.70&0.711&0.011 &0.018/0.036\\
       D &  0.70&0.708&0.008 &0.019/0.037\\
       E &  0.57&0.585&0.015 &0.018/0.036\\
        F & 0.96&0.911&0.049 &0.011/0.023\\
     G&    0.84&0.838&0.002 &0.014/0.029\\
        H & 0.80&0.825&0.025 &0.015/0.029\\
        I& 0.80&0.785&0.015 &0.016/0.032\\
      J&   0.63&0.624&0.006 &0.022/0.045\\
      R$^*$ & - & 0.701& - & 0.019/0.039\\
      \hline
        Average A-J & -& -& 0.0142&0.0166/0.0332\\
         \hline
        \hline
    \end{tabular}
      \caption{Overview on the results obtained with the calibrated CRN for regime 1. The average value is only a useful concept for the deviation between the mean prediction and experimental value and the (doubled) standard deviation $\sigma$. \\
        $^*$ unseen, no validation data available}
    \label{AP_Reg1_Pred_MeanExpDev2Stddev}
\end{table}


\clearpage
\subsection{Calibration of model parameters for PSR temperature in regime 2}

\begin{figure}[h]
    \centering
    \begin{subfigure}[b]{0.49\textwidth}
        \centering
        \includegraphics[width = \textwidth]{Figures/FiguresRD/FullPostPred_Reg2OP8.pdf}
        \caption{Full posterior prediction for OP I.}
    \end{subfigure}
        \begin{subfigure}[b]{0.49\textwidth}
        \centering
        \includegraphics[width = \textwidth]{Figures/FiguresRD/FullPostPred_Reg2OP9.pdf}
        \caption{Full posterior prediction for OP J.}
    \end{subfigure}
        \begin{subfigure}[b]{0.49\textwidth}
        \centering
        \includegraphics[width = \textwidth]{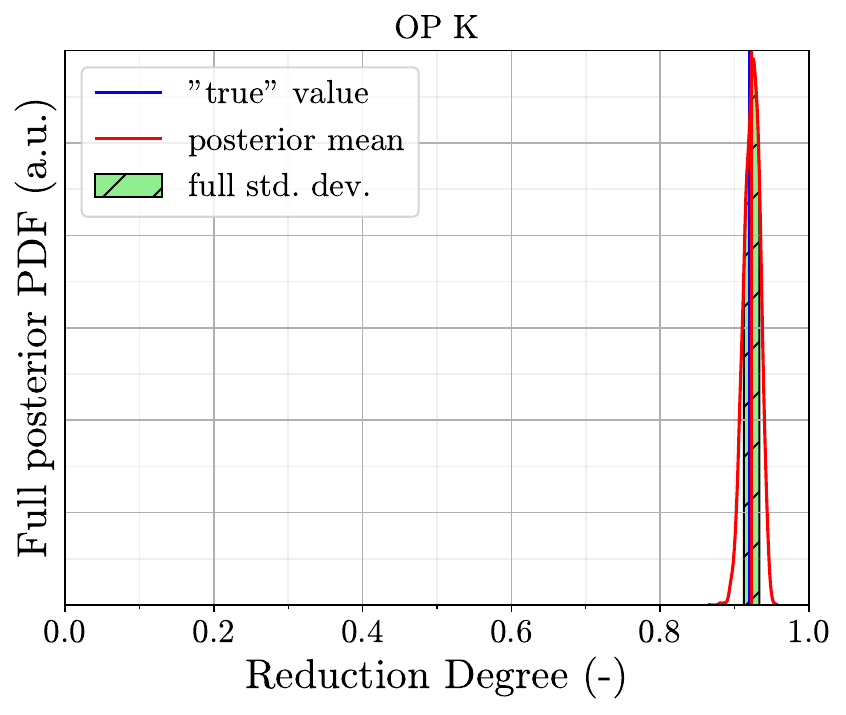}
        \caption{Full posterior prediction for OP K.}
    \end{subfigure}
        \begin{subfigure}[b]{0.49\textwidth}
        \centering
        \includegraphics[width = \textwidth]{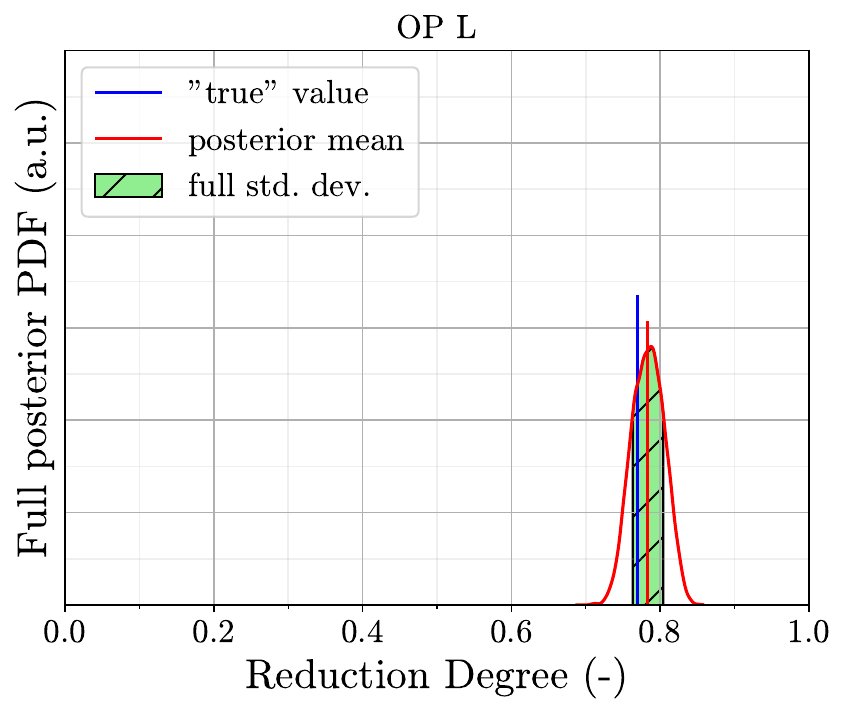}
        \caption{Full posterior prediction for OP L.}
    \end{subfigure}
            \begin{subfigure}[b]{0.49\textwidth}
        \centering
        \includegraphics[width = \textwidth]{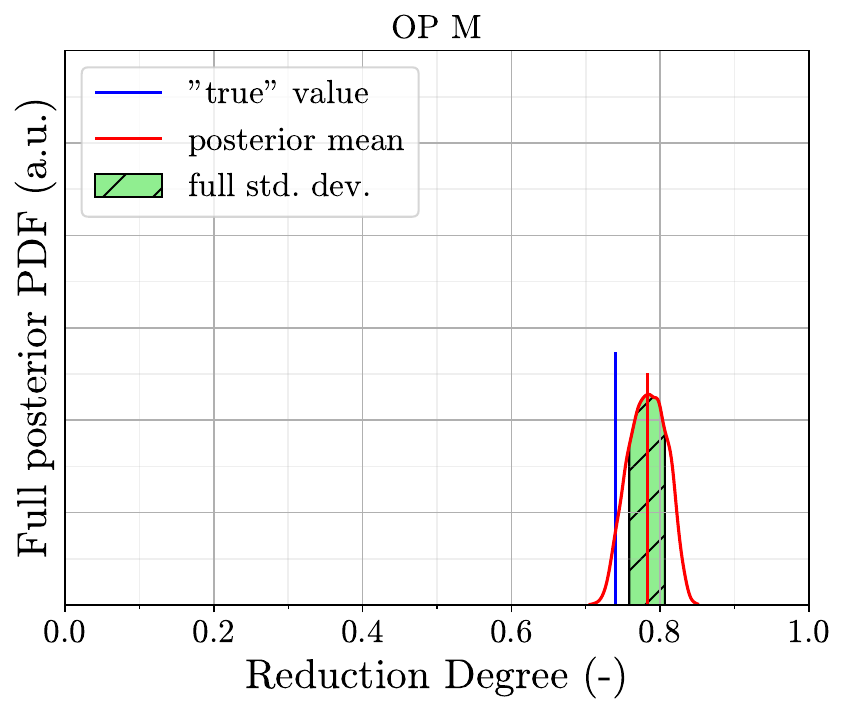}
        \caption{Full posterior prediction for OP M.}
    \end{subfigure}
            \begin{subfigure}[b]{0.49\textwidth}
        \centering
        \includegraphics[width = \textwidth]{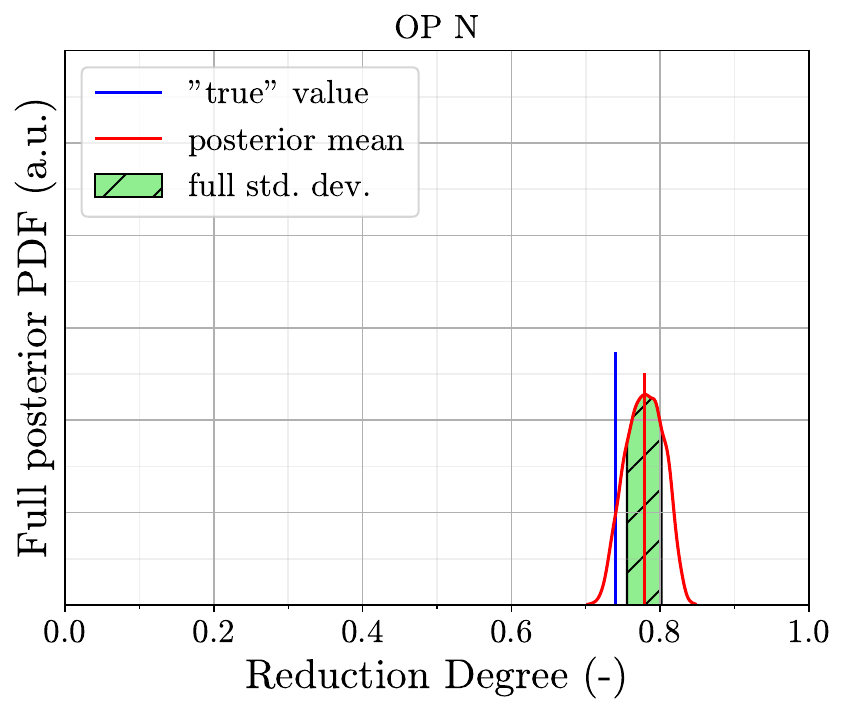}
        \caption{Full posterior prediction for OP N.}
    \end{subfigure}
    \caption{Overview of full posterior predictions for regime 2; OP I to N. }
        \label{AP_Reg2_ALLOPS_FullPostPredPDFsI-N}
    \end{figure}

\begin{figure}[h]
    \centering
                \begin{subfigure}[b]{0.49\textwidth}
        \centering
        \includegraphics[width = \textwidth]{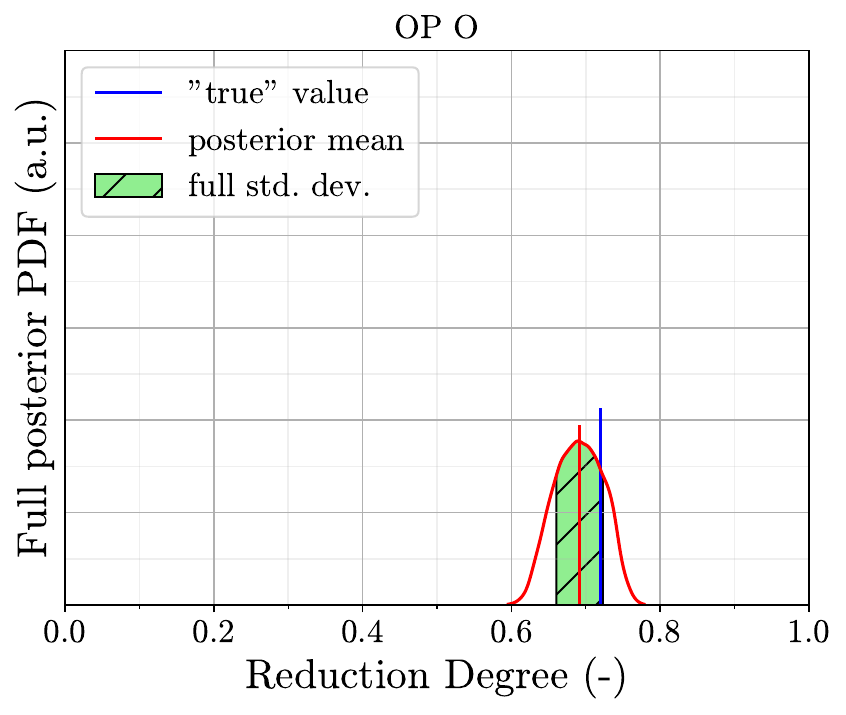}
        \caption{Full posterior prediction for OP O.}
    \end{subfigure}
                \begin{subfigure}[b]{0.49\textwidth}
        \centering
        \includegraphics[width = \textwidth]{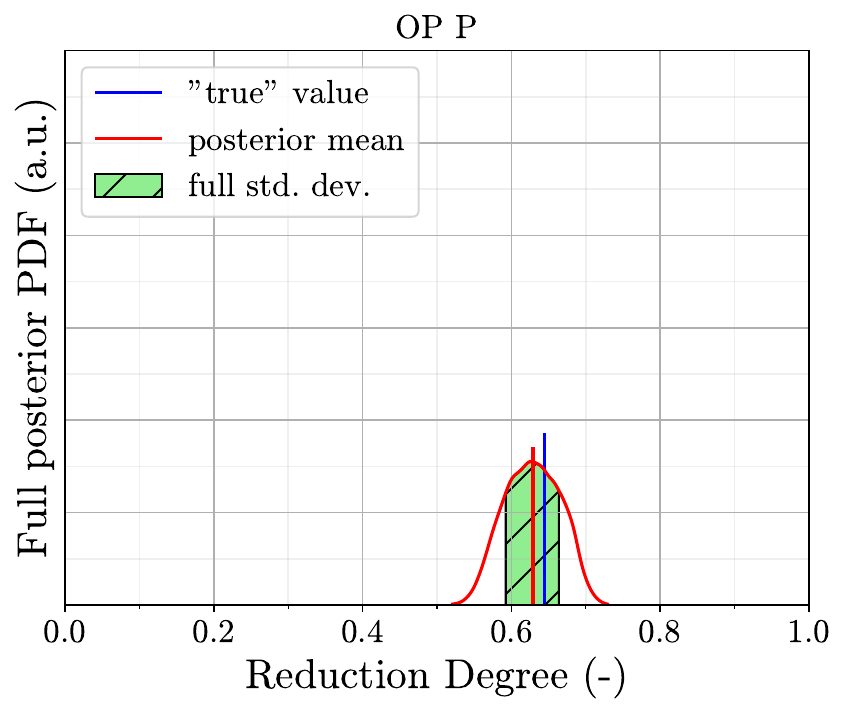}
        \caption{Full posterior prediction for OP P.}
    \end{subfigure}
                    \begin{subfigure}[t]{0.49\textwidth}
        \centering
        \includegraphics[width = \textwidth]{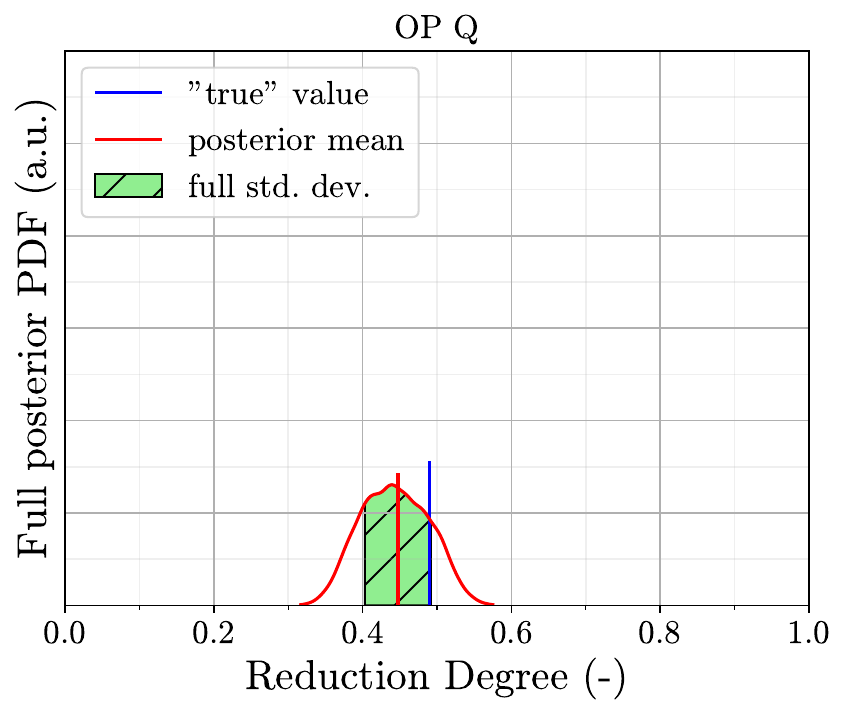}
        \caption{Full posterior prediction for OP Q.}
    \end{subfigure}
                    \begin{subfigure}[t]{0.49\textwidth}
        \centering
        \includegraphics[width = \textwidth]{Figures/FiguresRD/FullPostPred_Reg2OP17.pdf}
        \caption{Full posterior prediction for OP R (unseen, no validation data available).}
    \end{subfigure}
                \begin{subfigure}[b]{0.49\textwidth}
        \centering
        \includegraphics[width = \textwidth]{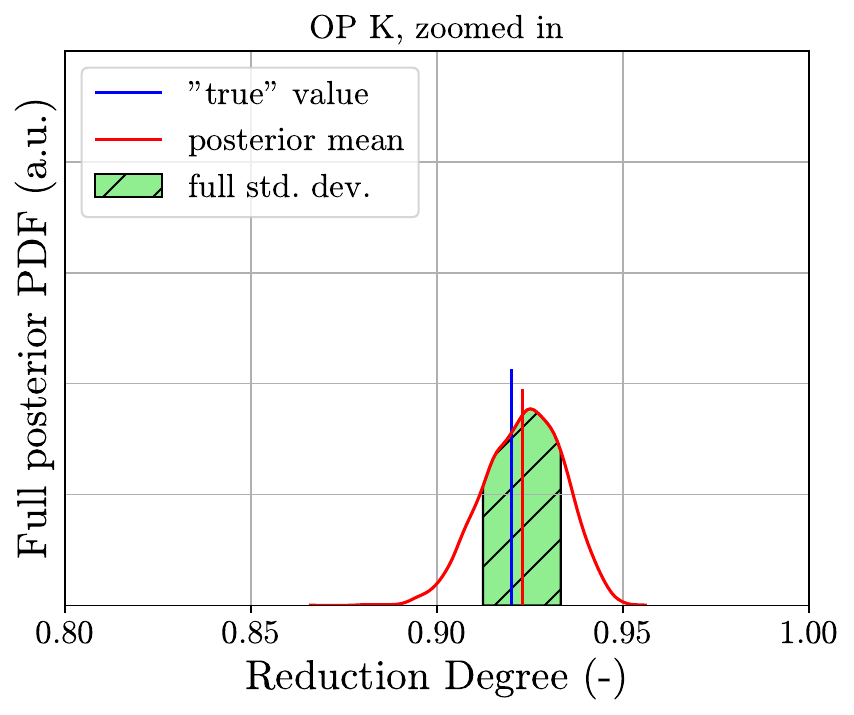}
        \caption{Full posterior prediction for OP K, zoomed in, with full y-axis.}
    \end{subfigure}
    \caption{Overview of full posterior predictions for regime 2; OP O-R; and K with a reduced x-axis and full y-axis for better visualization. }
    \label{AP_Reg2_ALLOPS_FullPostPredPDFsO-R}
\end{figure}

In Table~\ref{AP_Reg2_ALLOPS_Pred_MeanExpDev2Stddev}, the experimental reduction degrees reported by Sohn \cite{Sohn2023} are compared to the predictions of the CRN model developed within the present work, together with the deviation between the latter two values and width of the $1\sigma$-interval of the prediction, for the operation points of regime 2 for the second calibration. 
In Figures~\ref{AP_Reg2_ALLOPS_FullPostPredPDFsI-N} and~\ref{AP_Reg2_ALLOPS_FullPostPredPDFsO-R}, the full posterior PDFs of the predictions of the reduction degree obtained from the calibrated CRN model for regime 2 are shown. 


\begin{table}[h]
    \centering
    \begin{tabular}{ccccc}
    Operation point&Experimental value &Mean prediction & Deviation &$\sigma\ /\ 2\cdot \sigma$\\ 
    \hline
    \hline
       I   & 0.80&0.795&0.005 &0.020/0.040\\
       J & 0.63&0.600&0.03 &0.024/0.048\\
        K & 0.92&0.923&0.003 &0.011/0.021\\
       L &  0.77&0.783&0.013 &0.022/0.044\\
       M &  0.74&0.782&0.042 &0.025/0.049\\
        N & 0.74&0.779&0.039 &0.025/0.049\\
     O &    0.72&0.692&0.028 &0.032/0.064\\
        P & 0.645&0.629&0.016 &0.037/0.074\\
        Q& 0.49&0.448&0.042 &0.046/0.091\\
        R$^*$ & - &0.695& - &0.023/0.046\\
        \hline
        Average I-Q & -& -& 0.0242&0.027/0.054\\
                 \hline
        \hline
    \end{tabular}
            \caption{Overview on the results obtained with the calibrated CRN for regime 2. The average value is only a useful concept for the deviation between the mean prediction and experimental value and the (doubled) standard deviation $\sigma$. \\
        $^*$ unseen, no validation data available}
    \label{AP_Reg2_ALLOPS_Pred_MeanExpDev2Stddev}
\end{table}

\clearpage
\printbibliography
\end{document}